\documentstyle[12pt]{article}
\def\II{{I}}
\def\RR{{\bf R}}
\def\CC{{\bf C}}

\def\tr{{\rm tr\,}}
\def\Tr{{\rm Tr\,}}

\def\det{{\rm det\,}}

\def\End{{\rm End\,}}

\def\log{{\rm log\,}}

\def\rank{{\rm rank\,}}

\def\vol{{\rm vol\,}}

\def\be{\begin{equation}}
\def\ee{\end{equation}}
\def\bea{\begin{eqnarray}}
\def\eea{\end{eqnarray}}
\newtheorem{theorem}{Theorem}
\newtheorem{lemma}{Lemma}
\newtheorem{corollary}{Corollary}
\newtheorem{proposition}{Proposition}

\begin{document}

\begin{titlepage}
\null\vskip-3truecm\hspace*{7truecm}{\hrulefill}\par\vskip-4truemm\par
\hspace*{7truecm}{\hrulefill}\par\vskip5mm\par
\hspace*{7truecm}{{\large\sf University of Greifswald (October, 1997)}}
\vskip4mm\par\hspace*{7truecm}{\hrulefill}\par\vskip-4truemm\par
\hspace*{7truecm}{\hrulefill}\par
\bigskip\bigskip\hspace*{7truecm}
{\large\sf DSF 97/45}
\par
\vfill
\centerline{\LARGE\bf Gauge theories on manifolds with boundary}
\bigskip
\bigskip
\centerline{\Large\bf Ivan G. Avramidi
\footnote{On leave of absence from Research Institute for Physics, 
Rostov State University,  Stachki 194, 344104 Rostov-on-Don, Russia.}
\footnote{E-mail: iavramid@math.uiowa.edu}}
\bigskip
\centerline{\it Department of Mathematics, University of Greifswald}
\centerline{\it F.-L.-Jahnstr. 15a, D--17489 Greifswald, Germany}
\smallskip
\centerline{and}
\smallskip
\centerline{\it Department of Mathematics, The University of Iowa}
\centerline{\it 14 MacLean Hall, Iowa City, IA 52242-1419, USA}
\bigskip
\bigskip
\centerline{\Large\bf Giampiero Esposito}
\bigskip
\centerline{\it Istituto Nazionale di Fisica Nucleare, Sezione di Napoli}
\centerline{\it Complesso Universitario di M. S. Angelo, 
Edificio G, 80126 Napoli, Italy}
\smallskip
\centerline{and}
\smallskip
\centerline{\it Universit\`a di Napoli Federico II,
Dipartimento di Scienze Fisiche}
\centerline{\it Mostra d'Oltremare
Padiglione 19, 80125 Napoli, Italy}
\bigskip
\centerline{Revised on April 1998}
\medskip
\vfill

{\narrower\par
The boundary-value problem for Laplace-type 
operators acting on smooth sections of a vector bundle over a 
compact Riemannian manifold with generalized local boundary conditions 
including both normal and
tangential derivatives is studied.
The condition of strong ellipticity of this boundary-value problem
is formulated. The resolvent kernel and the
heat kernel in the leading approximation 
are explicitly constructed. As a result, the previous work in the
literature on heat-kernel asymptotics is shown to be a particular
case of a more general structure. For a bosonic gauge theory on a
compact Riemannian manifold with smooth boundary, the problem is
studied of obtaining a gauge-field operator of Laplace type,
jointly with local and gauge-invariant boundary conditions, 
which should lead to a strongly elliptic boundary-value problem.
The scheme is extended to fermionic gauge theories by means
of local and gauge-invariant projectors. After deriving a
general condition for the validity of strong ellipticity
for gauge theories, it is proved that for Euclidean
Yang--Mills theory and Rarita--Schwinger fields all the above
conditions can be satisfied. For Euclidean quantum gravity,
however, this property no longer holds, i.e. the corresponding 
boundary-value problem is not strongly elliptic.  
Some non-standard local formulae for the leading asymptotics of the 
heat-kernel diagonal are also obtained.
It is shown that, due to the absence of strong ellipticity,
the heat-kernel diagonal is non-integrable
near the boundary.

\par}
\vfill
\end{titlepage}

\section{Introduction}
\setcounter{equation}0

The consideration of boundary conditions in the formulation of
quantum field theories is crucial for at least two reasons:
\vskip 0.3cm
\noindent
(i) Boundary effects are necessary to obtain a complete prescription
for the quantization, unless one studies the idealized case where
no bounding surfaces occur. In particular, the functional-integral
approach relies heavily on a careful assignment of boundary data.
\vskip 0.3cm
\noindent
(ii) One may wonder whether the symmetries of the theory in the
absence of boundaries are preserved by their inclusion. Moreover,
one would like to know whether the application of a commonly 
accepted physical principle (e.g. the invariance under infinitesimal
diffeomorphisms on metric perturbations, in the case of
gravitation) is sufficient to determine completely the desired 
form of the boundary conditions, upon combination with other
well known mathematical properties.

The investigations carried out in our paper represent the attempt
to solve these problems by using the advanced tools of analysis
and geometry, with application to the operators of Laplace
or Dirac type. Indeed, the
differential operators of Laplace type are well known 
to play a crucial role in mathematical physics.
By choosing a suitable gauge it is almost
always possible to reduce the problem of evaluating the Green functions and
the effective action in quantum field theory to a calculation of Green
functions (and hence the resolvent) and functional 
determinants (or the $\zeta$-functions) of Laplace-type
operators. These objects are well defined,
strictly speaking, only for self-adjoint elliptic 
operators. Thus, on manifolds with boundary, 
one has to impose some boundary conditions which guarantee
the self-adjointness and the ellipticity of the 
Laplace-type operator. The simplest choice are 
the Dirichlet and Neumann boundary conditions
when the fields or their normal derivatives are set to zero on the boundary.
A slight modification of the Neumann boundary conditions are the so-called
Robin ones, when one sets to zero at the boundary 
a linear combination of the values of the
fields and their normal derivatives. 
An even more general scheme corresponds
to a mixed situation when some field components satisfy Dirichlet
conditions, and the remaining ones are subject to
Robin boundary conditions \cite{gilkey95,branson90}.
 
However, this is not the most general scheme, and one can 
define some {\it generalized } boundary conditions which 
are still local but include both normal and 
tangential derivatives of the fields
\cite{mcavity91}.
For example, by using linear covariant gauges in 
quantum gravity, one is led to impose boundary conditions that
involve the tangential derivatives as well, to
ensure that the whole set of boundary conditions on metric 
perturbations is invariant under infinitesimal diffeomorphisms
\cite{barvi87,avra96}. The same boundary conditions may be derived
by constructing a BRST charge and requiring BRST invariance
of the boundary conditions \cite{moss97}.  

In this paper we are going to study the
generalized boundary-value problem
for Laplace- and Dirac-type operators, the latter being relevant
for the analysis of fermionic models. 
Such a boundary-value problem, however, is not automatically elliptic.
Therefore, we find first an explicit criterion of ellipticity
(section 2.3). Then we construct the 
resolvent kernel and the heat kernel
in the leading approximation (section 3).
The application of this formalism (section 4) proves that the generalized
boundary-value problem is strongly elliptic for Euclidean
Yang--Mills theory (section 5) and
Rarita--Schwinger fields (section 6), but not for 
Euclidean quantum gravity (section 7).
The possible implications are discussed in section 8.

\section{Laplace-type and Dirac-type operators}
\setcounter{equation}0

Let $(M,g)$ be a smooth compact 
Riemannian manifold of dimension $m$ with smooth boundary, say $\partial M$.
Let $g$ be the positive-definite Riemannian metric on $M$ and $\hat g$ be  
the induced metric on $\partial M$.
Let $V$ be a (smooth) vector bundle over the manifold $M$ and
$C^\infty(V,M)$ be the space of smooth sections of the bundle $V$.
Let $V^*$ be the dual vector bundle and $E:\ V\to V^*$ be a Hermitian
non-degenerate metric,
$
E^{\dag}=E
$, 
that determines the Hermitian fibre scalar product in $V$.
Using the invariant Riemannian volume element 
$d\vol(x)$ on $M$ one defines a natural $L^2$ 
inner product $(,)$ in $C^\infty(V,M)$, and the Hilbert space
$L^2(V,M)$ as the completion of $C^\infty(V,M)$ in this norm.
Further, let $\nabla^V$ be the connection 
on the vector bundle $V$ compatible
with the metric $E$, $\tr_g=g\otimes 1$ be the 
contraction of sections of the bundle $T^*M \otimes 
T^*M \otimes V$ with the metric on the cotangent bundle
$T^*M $, and $Q$ be a smooth endomorphism of the bundle $V$, 
i.e. $Q\in \End(V)$,
satisfying the condition
\be
\bar Q\equiv E^{-1} Q^{\dag}E=Q .
\ee
Hereafter we call such endomorphisms self-adjoint.
We also use the notation $\bar A=E^{-1}A^{\dag}E$ for any endomorphisms
or operators.
Then a Laplace-type operator, or generalized Laplacian,
\be
F:\ C^\infty(V,M)\to C^\infty(V,M),
\ee
is a second-order differential operator defined by
\be
F \equiv -\tr_g\nabla^{T^*M \otimes V}\nabla^V+Q,
\label{4e}
\ee
where
\be
\nabla^{T^*M \otimes V}=\nabla^{T^*M }\otimes 1+1\otimes\nabla^{V},
\ee
and $\nabla^{T^*M }$ is the Levi-Civita connection on $M$.

Let $V$ be a Clifford bundle and $\Gamma:\ 
T^*M  \to \End(V)$, $\Gamma(\xi)=\Gamma^\mu\xi_\mu$, 
be the Clifford map satisfying
\be
\Gamma(\xi_1)\Gamma(\xi_2)+\Gamma(\xi_2)\Gamma(\xi_1)
=2g(\xi_1,\xi_2)\II ,
\label{2.5aaa}
\ee
for all $\xi_1,\xi_2\in T^*M $. Hereafter $\II $ denotes the identity 
endomorphism of the bundle $V$.
Let $\nabla^V$ be the Clifford connection compatible with
the Clifford map. Then a Dirac-type operator
\be
D:\ C^\infty(V,M)\to C^\infty(V,M),
\ee
is a first-order differential operator defined by
\cite{gilkey95, berline92}
\be
D \equiv i(\Gamma\nabla^{V}+S),
\label{4eee}
\ee
with $S\in\End(V)$.
Of course, the square of the Dirac operator is a Laplace-type operator.

\subsection{Geometry of boundary operators}

\subsubsection{Laplace-type operator}

For a Laplace-type operator we define the {\it boundary data}
by 
\be
\psi_F(\varphi)=\left(\matrix{\psi_0(\varphi)\cr
\psi_1(\varphi)\cr}\right),
\label{6e}
\ee
where
\be
\psi_0(\varphi)\equiv\varphi|_{\partial M},\qquad
\psi_1(\varphi)\equiv\nabla_N\varphi|_{\partial M}
\label{7e}
\ee
are the restrictions of the sections $\varphi\in C^\infty(V,M)$
and their normal derivatives, 
to the boundary (hereafter, $N$ is the inward-pointing 
unit normal vector field to the boundary).
To make the operator $F$ symmetric and elliptic (see sections 2.2 and 2.3),
one has to 
impose some conditions on the boundary data $\psi_F(\varphi)$.

In general, a $d$-graded vector bundle is a vector bundle
jointly with a fixed decomposition into 
$d$ sub-bundles \cite{gilkey95}.
In our problem, let the vector bundle $W_F$ over $\partial M$ 
be the bundle of the boundary data. 
$W_F$ consists of two copies of the restriction of $V$ to $\partial M$ and
inherits a natural grading
\cite{gilkey95}
\be
W_F=W_0\oplus W_1,
\ee
where $W_j$ represents normal derivatives of order $j$,
and, therefore,
\be
\dim W_F=2\dim V.
\label{2.12aa}
\ee
The bundles $W_0$ and $W_1$ have the same structure, 
and hence in the following they will
be often identified.
Let $W'_F=W'_0\oplus W'_1$ be an auxiliary graded
vector bundle over $\partial M$ such that
\be
\dim W'_F=\dim V.
\label{2.13aa}
\ee
Since the dimension of the bundle $W_F$ is twice as big as the dimension
of $W'_F$, $\dim W_F=2 \dim W'_F$, it is 
convenient to identify the bundle $W'_F$
with a sub-bundle of $W_F$ by means of a projection
\be
P_F:\ W_F\to W'_F, \qquad P_F^2=P_F.
\ee
In other words, sections of the bundle $W'_F$ have the form $P_F\chi$, with
$\chi$ being a section of the bundle $W_F$.
Note that the rank of the projector $P_F$ is equal to the dimension of the
bundle $V$,
\be
\rank P_F=\rank (\II_W-P_F)=\dim V,
\label{14ee}
\ee
where $\II_W$ is the identity endomorphism of the bundle $W_F$.

Let $B_F: C^\infty(W_F,\partial M)\to 
C^\infty(W'_F,\partial M)$ be a tangential differential operator
on $\partial M$. The boundary conditions then read
\be
B_F\psi_F(\varphi)=0.
\label{14e}
\ee
The boundary operator $B_F$ is not arbitrary but should satisfy some 
conditions to make the operator $F$ self-adjoint and elliptic. These conditions
are formulated in the subsects. 2.2 and 2.3.
The boundary operator $B_F$ can be 
then presented in the form
\be
B_F=P_F L,
\ee
where 
$L$ is a {\it non-singular} tangential operator 
\be
L: C^\infty(W_F,\partial M)\to C^\infty(W_F,\partial M),
\ee
meaning that there is a well defined inverse operator $L^{-1}$.
In other words, the boundary conditions mean
\be
\psi_{F}(\varphi)=K_F\chi \qquad {\rm with \ arbitrary }
\ \chi\in C^\infty(W_F,\partial M), 
\ee
where
\be
K_F=L^{-1}(\II_W-P_F):\ 
C^\infty(W_F,\partial M)\to C^\infty(W_F,\partial M).
\ee

Now let $\Pi$ be a self-adjoint projector acting 
on $W_0$ and $W_{1}$,
\be
\Pi:\ W_0\to W_0,
\qquad
\Pi:\ W_{1} \to W_{1}
\ee
\be
\Pi^2=\Pi,
\qquad
{\bar \Pi} \equiv E^{-1} \Pi^{\dagger} E =\Pi .
\ee
In our analysis, we will consider the projector $P_F$ and 
the operator $L$ of the form
\be
P_F=\Pi\oplus(\II-\Pi)=\left(\matrix{
\Pi&0\cr
0&(\II-\Pi)\cr}\right),
\ee
\be
L=\left(\matrix{\II&0\cr
\Lambda& \II \cr}\right),
\ee
where $\Lambda$ is a tangential 
differential operator on $\partial M$,
$
\Lambda:\ C^\infty(W_0,\partial M)\to C^\infty(W_0,\partial M),
$
satisfying the conditions
\be
\Pi\Lambda=\Lambda\Pi=0.
\label{24e}
\ee 
Of course, $\rank P_F=\rank\Pi+\rank(\II-\Pi)=\dim V$ as needed.
Hence we obtain
\be
B_F=\left(\matrix{\Pi&0\cr
\Lambda& \II_{W}-\Pi\cr}\right),\qquad
K_F= \II_W-B_{F}=\left(\matrix{\II_{W}-\Pi&0\cr
-\Lambda&\Pi\cr}\right).
\label{25e}
\ee
Note that $B_F$ and $K_F$ are complementary projectors
\be
B_F^2=B_F, \qquad K_F^2=K_F, \qquad B_FK_F=K_FB_F=0.
\ee
Moreover, by virtue of the property (\ref{14ee}), both the operators
$B_F$ and $K_F$ do not vanish.

\subsubsection{Dirac-type operator}

For a Dirac-type operator (see (\ref{4eee})) 
the normal derivatives are {\it not} included
in the boundary data $\psi_D(\varphi)$, and such 
boundary data consist only of
the restriction $\psi_D(\varphi)=\psi_0(\varphi)$ (see (\ref{7e}))
of sections $\varphi\in C^\infty(V,M)$ to the boundary.
Therefore, the bundle of the boundary data $W_D$ 
is just the restriction $W_0$ of the bundle $V$
and, similarly, the auxiliary vector bundle $W'_D=W'_0$.
The dimensions of these bundles are 
\be
\dim W_D=2\dim W'_D=\dim V . 
\ee
This should be compared with (\ref{2.12aa}) and (\ref{2.13aa}).
As above, $W'_D$ can be identified with a sub-bundle of $W_D$
by means of a projection
\be
P_D:\ W_D\to W'_D,
\ee
but now the dimension of the projector $P_D$ is 
equal to {\it half} the dimension
of the bundle $V$:
\be
\rank P_D={1\over 2}\dim V.
\ee
The boundary conditions for the operator $D$ read
\be
B_D\psi_0(\varphi)=0,
\label{2.31aa}
\ee
where now the operator $B_D$ is not a tangential differential
operator but just a projector $P_D$:
\be
B_D=P_D, \qquad K_D=\II -P_{D}.
\label{2.32aa}
\ee
It is crucial to use projectors to obtain a well posed 
boundary-value problem of local nature for Dirac-type operators,
since any attempt to fix the whole fermionic field at the
boundary would lead to an over-determined problem.

\subsection{Symmetry}

\subsubsection{Laplace-type operator}

The Laplace-type operator $F$ is {\it formally
self-adjoint}, which means that it is symmetric on any 
section of the bundle $V$
with compact support {\it disjoint} from the boundary $\partial M$, 
i.e.
\be
I_F(\varphi_1,\varphi_2)\equiv 
(F\varphi_1,\varphi_2)-(\varphi_1,F\varphi_2)=0,
\qquad {\rm for}\
\varphi_1,\varphi_2\in C^\infty_0(V,M).
\label{27}
\ee
However, a formally self-adjoint operator is not necessarily
self-adjoint. It is {\it essentially self-adjoint} if
its closure is self-adjoint. This implies that 
the operator $F$ is such that:
i) it is symmetric on any smooth section satisfying the boundary conditions
(\ref{14e}), i.e.
\be
I_{F}(\varphi_1,\varphi_2)=0,
\qquad
{\rm for }\ \ \varphi_{1}, \varphi_{2} \in C^\infty(V,M):
\qquad B_F\psi(\varphi_1)=B_F\psi(\varphi_2)=0,
\ee
and ii) there exists a unique self-adjoint extension of $F$.
The property ii) can be proved by studying deficiency
indices \cite{resi75},
but is not the object of our investigation.

The antisymmetric 
bilinear form $I_F(\varphi_1,\varphi_2)$ 
depends on the boundary data.
Integrating by parts, it is not difficult to obtain
\be
I_F(\varphi_1,\varphi_2)=<\psi(\varphi_{1}),J_F\psi(\varphi_{2})> ,
\label{33a}
\ee
where
\be
J_F=\left(\matrix{0&\II\cr
-\II&0\cr}\right).
\ee
Here $<,>$ denotes the $L^2$ inner product in $C^\infty(W,\partial M)$ 
determined by the restriction
of the fibre metric to the boundary.

To ensure the symmetry, we have to fix the
boundary operators $B_F$ so as to make this form 
{\it identically} zero. The condition for that reads
\be
\bar K_FJ_FK_F= 0.
\label{22e}
\ee
By using the general form (\ref{25e}) of the operator $K_F$, 
we find herefrom that 
this is equivalent to the condition of symmetry of the
operator $\Lambda$ 
\be
<\Lambda\varphi_1,\varphi_2>=<\varphi_1,\Lambda\varphi_2>,\qquad
{\rm for}\ 
\varphi_1,\varphi_2\in C^\infty(W_0,\partial M),
\label{24ee}
\ee

Thus, we have proven the following result:
\begin{theorem}
The Laplace-type operator $F$ (\ref{4e}) endowed with the 
the boundary conditions (\ref{14e}), with the boundary operator $B$ 
given by (\ref{25e}), and any symmetric operator
$\Lambda$ satisfying the conditions (\ref{24e}) and (\ref{24ee}), 
is symmetric.
\end{theorem}

\subsubsection{Dirac-type operator}

In the case of Dirac-type operators there are two essentially 
different cases. The point is that, in general, there exist two 
{\it different} representations of the 
Clifford algebra satisfying
\be
\bar \Gamma(\xi)=-\varepsilon\Gamma(\xi),\qquad 
\varepsilon=\pm 1.
\label{2.39bbb}
\ee
In odd dimension $m$ there is only one possibility,
$\varepsilon=-1$, corresponding to self-adjoint Dirac matrices, 
whereas for even dimension $m$ they can be either 
self-adjoint or anti-self-adjoint.
By requiring the endomorphism $S$ to satisfy the condition
\be
\bar S=\varepsilon S,
\ee
we see that the Dirac-type operator (\ref{4eee}) is formally 
self-adjoint for $\varepsilon=-1$ and anti-self-adjoint
for $\varepsilon=1$:
\be
I_D(\varphi_1,\varphi_2)\equiv 
(D\varphi_1,\varphi_2)+\varepsilon(\varphi_1,D\varphi_2)=0,
\qquad
\varphi_1,\varphi_2\in C^\infty_0(V,M).
\label{27bbb}
\ee
In complete analogy with the above, we easily find the condition
for the Dirac-type operator to be (anti)-symmetric,
\be
I_D(\varphi_1,\varphi_2)=<\psi_0(\varphi_{1}),
J_D\psi_0(\varphi_{2})>=0, 
\label{33aa}
\ee
where
\be
J_D=i\varepsilon\Gamma(N)=i\varepsilon\Gamma^\mu N_\mu,
\ee
for any $\varphi_1, \varphi_2 \in C^\infty(V,M)$ satisfying the boundary 
conditions (\ref{2.31aa}). This leads to
a condition on the boundary operator,
\be
\bar K_DJ_DK_D= 0,
\ee
wherefrom, by using (\ref{2.32aa}), 
we obtain a condition for the projector $P_{D}$,
\be
(\II-\bar P_D)\Gamma(N)(\II-P_D)=0.
\label{2.42aa}
\ee
Hence we get a sufficient condition on the boundary projector, 
\be
\bar P_D=\Gamma(N)(\II-P_D)\Gamma(N)^{-1}.
\ee
This means that $P_D$ can be expressed in the form
\be
P_D={1\over 2}(\II+\eta ),
\ee
where $\eta $ satisfies the conditions
\be
\eta ^2=\II, \qquad \bar \eta \Gamma(N)+\Gamma(N)\eta =0.
\label{2.48bbb}
\ee
Thus, there are two cases: i) $\eta $ is 
anti-self-adjoint and commutes with $\Gamma(N)$,
ii) $\eta $ is self-adjoint and anti-commutes with $\Gamma(N)$.
This leads to two particular solutions,
\be
\eta =\pm \Gamma(N), \qquad {\rm for}\  
\varepsilon=1, \qquad {\rm and\ even}\ \ m,
\label{2.49bb}
\ee
and
\be
\eta ={1\over |u|}\Gamma(u), \qquad {\rm for}\ \varepsilon=-1\qquad
{\rm and\ any}\ \ m,
\label{2.50bb}
\ee
with some cotangent vector $u\in T^*\partial M $ on the boundary.

In {\it even} dimension $m$ 
there exists another very simple solution,
\be
\eta =\pm C, \qquad {\rm for\ even}\ m, \qquad  \varepsilon=\pm 1, 
\label{2.51bb}
\ee
where $C$ is the {\it chirality} operator defined with the help of
an orthonormal basis $e^a$ in $T^*M $,
\be
C=i^{m/2}\Gamma(e^1)\cdots\Gamma(e^m).
\label{2.49aaa}
\ee
One easily finds 
\be
C^2=\II, \qquad \bar C=C,
\ee
\be
C\Gamma(\xi)+\Gamma(\xi)C=0,
\label{2.52aaa}
\ee
for any $\xi\in T^*M $, so that the conditions (\ref{2.48bbb}) are satisfied.

Thus, we have 
\begin{theorem}
The Dirac-type operator $D=i(\Gamma\nabla+S)$ 
with $\Gamma$ and $S$ satisfying the conditions
$\bar\Gamma=-\varepsilon\Gamma$, $\bar S={\varepsilon}S$,
endowed with the boundary conditions $P_D\varphi|_{\partial M}=0$,
is anti-symmetric for $\varepsilon=-1$ and symmetric for 
$\varepsilon=1$ provided that the boundary projector $P_D$ satisfies
the condition $(\II-\bar P_D)\Gamma(N)(\II-P_D)=0$. 
Admissible boundary projectors satisfying this criterion
are: for $\varepsilon=1$
\be
P_D={1\over 2}\left[\II \pm \Gamma(N)\right],
\label{2.57aab}
\ee
and, for $\varepsilon=-1$,
\be
P_D(u)={1\over 2}\left[\II+{1\over|u|}\Gamma(u)\right],
\label{2.56aa}
\ee
where $u \in T^*\partial M $. In the case of an even-dimensional
manifold $M$, the boundary projector 
\be
P_D={1\over 2}\left(\II \pm C\right),
\label{2.57aa}
\ee
$C$ being the chirality operator, is also admissible.

\end{theorem}

\subsection{Strong ellipticity}

\subsubsection{Laplace-type operator}

Now we are going to study the ellipticity of the 
boundary-value problem defined by the
boundary operator (\ref{25e}). First of all we 
fix the notation. By using the inward geodesic flow, 
we identify a narrow neighbourhood of the boundary $\partial M$ with a part of 
$\partial M\times \RR_+$ and define a split of the cotangent bundle 
$T^*M =T^*\partial M \oplus T^*(\RR)$. 
Let $\hat x=(\hat x^i)$, $(i=1,2,\dots,m-1)$, be
the local coordinates on $\partial M$ and $r$ be the normal geodesic distance
to the boundary, so that $N=\partial_r=
\partial/\partial r$ is the inward unit 
normal on $\partial M$. Near $\partial M$ 
we choose the local coordinates $x=(x^\mu)=(\hat x,r)$, 
$(\mu=1,2,\dots,m)$ and the split $\xi=(\xi_\mu)=(\zeta,\omega) 
\in T^*M $, where
$\zeta=(\zeta_j) \in T^*\partial M $ and $\omega\in \RR$.
With our notation, Greek indices run 
from 1 through $m$ and lower case Latin indices run from 1
through $m-1$.

Our presentation differs from
the one in \cite{avresp97b}, since we always work 
with self-adjoint projectors.
In this paper we are interested in the so-called generalized 
boundary conditions, when $\Lambda$ is a 
{\it first-order} tangential differential operator
acting on sections of the vector bundle $W_0$ over $\partial M$.
Any formally self-adjoint operator of first order satisfying the 
conditions (\ref{24e}) can be put in the form
(hereafter, ${\hat \nabla}_{i}$ denotes $(m-1)$-dimensional
covariant differentiation tangentially, defined in 
ref. \cite{gilkey95})
\be
\Lambda=(\II-\Pi)\left\{
{1\over 2}(\Gamma^i\hat\nabla_i+\hat\nabla_i\Gamma^i)
+S\right\}(\II-\Pi),
\label{55}
\ee
where $\Gamma^i\in C^\infty(T\partial M 
\otimes {\rm End}\,(W_0),\partial M)$ are some
endomorphism-valued vector fields on $\partial M$, and $S$ is
some endomorphism of the vector bundle $W_0$, satisfying the conditions
\be
\bar\Gamma^i=-\Gamma^i, \qquad \bar S=S,
\label{34e}
\ee
\be
\Pi\Gamma^i=\Gamma^i\Pi=\Pi S=S\Pi=0.
\label{55d}
\ee
Now we are going to determine under which conditions the boundary-value 
problem is strongly elliptic \cite{gilkey95}.

First of all, the leading symbol of the operator 
$F$ should be elliptic in the interior of $M$.
The leading symbol of the operator $F$ reads
\be
\sigma_L(F;x,\xi)=|\xi|^2\cdot \II\equiv 
g^{\mu\nu}(x)\xi_\mu\xi_\nu \cdot \II ,
\ee
where $\xi\in T^*M $ is a cotangent vector.
Of course, for a positive-definite 
non-singular metric the leading symbol is
non-degenerate for $\xi\ne 0$. Moreover, for a 
complex $\lambda$ which does not
lie on the positive real axis, $\lambda\in {\bf C}-{\bf R}_+$
(${\bf R}_+$ being the set of positive numbers),
\be
\det (\sigma_L(F;x,\xi)-\lambda)=(|\xi|^2-\lambda)^{\dim V}\ne 0.
\ee
This equals zero only for $\xi=\lambda=0$.
Thus, the leading symbol of the operator $F$ is {\it elliptic}.

Second, the so-called {\it strong ellipticity condition} should be satisfied
\cite{gilkey95,gilkeysmith83b}.
As we already noted above, there is a natural grading in the vector bundles
$W_F$ and $W'_F$ which reflects simply the number of normal derivatives
of a section of the bundle \cite{gilkey95}.
The boundary operator $B_F$ (\ref{25e}) 
is said to have the {\it graded order} $0$.
Its {\it graded leading symbol} is defined by
\cite{gilkey95,gilkeysmith83b}
\be
\sigma_{g}(B_{F}) \equiv \left(\matrix{\Pi&0\cr
iT& (\II-\Pi)}\right),
\ee
where, by virtue of (\ref{55}),
\be
T=-i\sigma_L(\Lambda)=\Gamma^{j}\zeta_{j} ,
\label{83}
\ee
$\zeta\in T^*\partial M $ being a cotangent vector on the boundary.
By virtue of (\ref{34e}) the matrix $T$ is anti-self-adjoint,
\be
\bar T=-T,
\ee
and satisfies the conditions
\be
\Pi T=T\Pi=0.
\label{2.41ee}
\ee
To define the strong ellipticity condition we
take the leading symbol 
$\sigma_L(F;\hat x, r, \zeta, \omega)$ of the operator $F$, 
substitute $r=0$ and $\omega\to -i\partial_{r}$ and
consider the following ordinary differential equation 
for a $\varphi\in C^\infty(V,\partial M\times {\bf R}_+)$:
\be
\left[\sigma_{L}(F;\hat x, 0, 
\zeta, -i\partial_{r})-\lambda\cdot I\right]\varphi(r)=0,
\label{80}
\ee
with an asymptotic condition
\be
\lim_{r\to\infty}\varphi(r)=0,
\label{267}
\ee
where $\zeta\in T^*\partial M $, $\lambda\in {\bf C}-{\bf R}_+$ 
is a complex number which 
does not lie on the positive real axis, and $(\zeta,\lambda)\ne (0,0)$. 

The boundary-value
problem $(F,B)$ is said to be {\it strongly elliptic} 
\cite[p.415]{gilkeysmith83b}
with respect to the cone $\CC-\RR_{+}$ if for every
$\zeta\in T^*\partial M $, $\lambda \in \CC-\RR_+$, $(\zeta,\lambda)\ne (0,0)$, 
and $\psi'\in C^\infty(W',\partial M)$  there is a
{\it unique} solution $\varphi$ to the equation (\ref{80})
subject to the condition (\ref{267}) and satisfying
\be
\sigma_g(B_{F})({\hat x},\zeta)\psi(\varphi)=\psi' ,
\ee
with $\psi(\varphi)\in C^\infty(W,\partial M)$ being the boundary 
data defined by (\ref{6e}) and (\ref{7e}).

A purely algebraic formulation of strong
ellipticity can also be given, following Gilkey and Smith 
\cite{gilkeysmith83b}.
For this purpose, let us denote by $W_F^{\pm}(\zeta,\lambda)$ the
subsets of $W_{F}$ corresponding to boundary data of solutions
of Eq. (\ref{80}) vanishing as $r \rightarrow \pm \infty$.
Decompose $\sigma_L(F;\hat x,0,\zeta,\omega)=p_0\omega^2
+p_{1}\omega+p_{2}$,
where $p_j$ is homogeneous of order $j$ in $\zeta$. 
Then the differential equation (\ref{80})
can be rewritten in the form of a first-order system
\be
-i\Bigr[\partial_{r}+\tau(\zeta,\lambda)\Bigr]
\left(\matrix{\varphi\cr
\varphi_1
}\right)=0,
\ee
where
\be
\tau=i
\left(\matrix{
0&-1\cr
p_0^{-1}(p_2-\lambda)&p_0^{-1}p_1
\cr}
\right).
\ee
Herefrom one sees that $\tau$ does not have any 
purely-imaginary eigenvalues for 
$(\zeta,\lambda)\ne (0,0)$ if $(F,B)$ is strongly elliptic.
It is then possible to re-express the strong ellipticity condition 
by saying that 
\be
\sigma_{g}(B_{F})({\hat x},\zeta): \ W_F^{+}(\zeta,\lambda)
\rightarrow W_{F}'
\label{2.71nnn}
\ee
should be an {\it isomorphism}, for $(\zeta,\lambda) \not
= (0,0), \zeta \in T^{*}({\partial M}), \lambda \in
\CC-\RR_{+}$ \cite[p.416]{gilkeysmith83b}.

For a Laplace-type operator the equation (\ref{80}) takes the form
\be
\left[-\partial_r^2+|\zeta|^2 -\lambda\right]\varphi(r)=0,
\label{80a}
\ee
where $|\zeta|^2 \equiv \hat g^{ij}(\hat x)\zeta_i\zeta_j$.
The general solution satisfying the decay condition at infinity,
$r\to\infty$, reads
\be
\varphi(r)=\chi\exp(-\mu r),
\ee
where $\mu \equiv \sqrt{|\zeta|^2 -\lambda}$. 
Since $(\zeta,\lambda)\ne (0,0)$ and 
$\lambda \in {\bf C}-{\bf R}_+$, one can always choose ${\rm Re}\, \mu >0$.

The boundary data are now
\be
\psi_F(\varphi)=\left(
\matrix{\chi\cr
-\mu \chi}\right) .
\ee
Thus, the question of strong ellipticity for 
Laplace-type operators is reduced
to the invertibility of the equations 
\be
\left(\matrix{\Pi&0\cr
iT& (\II-\Pi)}\right)
\left(
\matrix{\chi\cr
-\mu \chi}\right)
=\left(
\matrix{\psi'_0\cr
\psi'_1}\right)
\label{48e}
\ee
for arbitrary $\psi'_0\in C^\infty(W'_0,\partial M), \psi'_1\in C^\infty(W'_1,\partial M)$.
This is obviously equivalent to the algebraic criterion (\ref{2.71nnn}).
The eq. (\ref{48e}) can be transformed into
(cf. \cite{avresp97b})
\bea
&&\Pi\chi=\psi'_{0} ,
\label{50aa}\\
&&
(\mu\II -iT)\chi=\mu\psi'_{0}-\psi'_1.
\label{50e}
\eea
Remember that $\psi'_F=P_F\tilde\psi_F$ with some 
$\tilde\psi_F\in C^\infty(W_F,\partial M)$. Therefore,
$(1-\Pi)\psi'_0=\Pi\psi'_1=0$ and 
the first equation follows from the second one.
Therefore, if the equation (\ref{50e}) has a 
{\it unique solution for any} $\psi'_{0}$ 
and $\psi'_1$,
then the boundary-value problem is strongly elliptic.
A {\it necessary and sufficient} condition to achieve this is expressed  
by the non-degeneracy of the matrix $[\mu\II-iT]$, i.e.
\be
\det[\mu\II-iT]\ne 0 ,
\label{52e}
\ee
for any $(\zeta,\lambda)\ne (0,0)$ and $\lambda\in {\bf C}-{\bf R}_+$.
In this case the solution of eq. (\ref{50e}) reads
\be
\chi=(\mu\II -iT)^{-1}\left[\mu\psi'_0-\psi'_1\right].
\label{53e}
\ee

Since the matrix $iT$ is self-adjoint, it has only real eigenvalues, in 
other words the eigenvalues of $T^2$ are real and {\it negative},
$T^2\le 0$.
It is clear that, for any non-real $\lambda\in 
{\bf C}-{\bf R}$, $\mu=\sqrt{|\zeta|^2-\lambda}$ 
is complex and, therefore, the matrix $[\mu\II-iT]$ is non-degenerate.
For real negative $\lambda$, 
$\mu$ is real and we have $\mu>|\zeta|$. 
Thus, the condition (\ref{52e}) means that the
matrix $|\zeta| \II -iT$ is positive-definite,
\be
|\zeta|\II-iT>0.
\label{52eaa}
\ee
A {\it sufficient} condition for that reads
\be
|\zeta|^2\II+T^2>0.
\label{52eaaa}
\ee
On the other hand, there holds, of course, $|\zeta|^2\II+T^2\le |\zeta|^2$.
The eq. (\ref{52eaaa}) means that the absolute values of 
all eigenvalues of the matrix $(iT)$,
both positive and negative, are smaller than $|\zeta|$, 
whereas (\ref{52eaa}) means that only the 
positive eigenvalues are smaller than
$|\zeta|$, but says nothing about the negative ones.
A similar inequality has been derived in \cite{dowker97}, but in that
case the boundary operator does not include the effect of $\Pi$,
following \cite{mcavity91,avresp97}.

This proves the following theorem:
\begin{theorem}
Let $F$ be a Laplace-type operator defined by
(\ref{4e}), and $B_F$ the generalized boundary 
operator given by (\ref{25e}) with
the operator $\Lambda$ defined by (\ref{55}). 
Let $\zeta\in T^*\partial M $ be 
a cotangent vector on the boundary and
$T \equiv \Gamma^{j}\zeta_{j}$. The boundary-value problem 
$(F,B_F)$ is strongly elliptic with respect to
$\CC-\RR_{+}$ if and only if for any non-vanishing $\zeta\ne 0$ 
the matrix $|\zeta| \II -iT$ is positive-definite, i.e. $|\zeta|\II-iT>0$.
\end{theorem}

\subsubsection{Dirac-type operator}

The question of strong ellipticity for boundary-value problems
involving Dirac-type operators can also be studied. As is
well known, the leading symbol of a Dirac-type operator reads
\be
\sigma_L(D;\xi)=-\Gamma(\xi).
\ee
Since the square of a Dirac-type operator is a Laplace-type operator,
it is clear that the leading symbol $\sigma_L(D;x,\xi)$ is 
non-degenerate in the interior of $M$ for $\xi\ne 0$.
Moreover, for a complex $\lambda$ one finds
\be
\left[\sigma_L(D;\xi)-\lambda\II\right]^{-1}=
\left[\sigma_L(D;\xi)+\lambda\II\right]{1\over |\xi|^2-\lambda^2}.
\ee
Therefore, $[\sigma_L(D;\xi)-\lambda\II]$ is non-degenerate
when $|\xi|^2-\lambda^2\ne 0$. But this vanishes only 
for ${\rm Im}\,\lambda=0$ and
${\rm Re}\,\lambda=\pm |\xi|$ and {\it arbitrary} $\xi$.
Thus, for $(\xi,\lambda)\ne (0,0)$ and 
$\lambda \in {\bf C}-{\bf R}_{+}-{\bf R}_{-}$, 
$[\sigma_{L}(D;\xi)-\lambda\II]$ is non-degenerate.

The boundary-value problem
for a Dirac-type operator $D$ is strongly elliptic if 
there exists a unique solution of the equation
\be
\left[i\Gamma(N)\partial_r-\Gamma(\zeta)
-\lambda\II\right]\varphi(r)=0,
\label{2.74aa}
\ee
for any $(\zeta,\lambda)\ne (0,0)$, $\lambda\notin 
{\bf R}_{+} \cup {\bf R}_{-}$, such that
\be
\lim_{r\to\infty}\varphi(r)=0,
\ee
\be
P_D\psi_D(\varphi)=\psi'_{D},
\ee
for any $\psi'_D\in W'_D$.

The general solution of (\ref{2.74aa}) satisfying 
the decay condition at infinity reads
\be
\varphi(r)=\chi\exp(-\mu r),
\ee
where now $\mu \equiv \sqrt{|\zeta|^2-\lambda^2}$.
Again, since $(\zeta,\lambda)\ne (0,0)$, 
$\lambda\in {\bf C}-{\bf R}_+-{\bf R}_-$, 
the root can be always defined by ${\rm Re}\,\mu>0$.
The constant prefactor $\chi$ should satisfy the equation
\be
(X-\lambda\II)\chi=0,
\label{2.87aa}
\ee
where
\be
X \equiv \sigma_L(D;i\mu N,\zeta)=-i\mu\Gamma(N)-\Gamma(\zeta),
\ee
and the boundary condition
\be
P_D\chi=\psi'_{D},
\label{2.88aa}
\ee
with some $\psi'_D=P_D\tilde\psi_0$, $\tilde\psi_0\in W_0$.
The eqs. (\ref{2.87aa}), (\ref{2.88aa}) are reduced to
\bea
&&\chi=\psi'_{D}+\chi_{-}, \\
&&\left[X+\lambda(2P_{D}-\II)\right]\chi_-=-(X-\lambda\II)\psi'_D.
\label{2.90}
\eea
Note that, since $P_{D}\chi_{-}$ vanishes, the coefficient
of $P_{D}$ in Eq. (\ref{2.90}) is arbitrary, and is set 
equal to 2 for convenience. 
Thus, the question of strong ellipticity for a Dirac-type
operator is reduced to the matrix $[X+\lambda(2P_{D}-\II)]$ 
being non-degenerate,
\be
\det\left[X+\lambda(2P_{D}-\II)\right]\ne 0,
\ee
for any $(\zeta,\lambda)\ne (0,0)$, 
$\lambda\in {\bf C}-{\bf R}_{+}-{\bf R}_{-}$.
If this is satisfied, then the solution reads
\be
\chi=2\lambda\left[X+\lambda(2P_D-\II)\right]^{-1}\psi'_D.
\ee
Let us set $(2P_D-\II)\equiv \eta $ and 
let us compute the square of the matrix $[X+\lambda \eta ]$.
For the boundary projectors defined by (\ref{2.57aab})--(\ref{2.57aa})
the matrix $\eta $ is either $\eta =\Gamma(N)$, 
$\eta =\Gamma(u)/|u|$ with some $u\in T^*\partial M $
or (in even dimension $m$ only) is exactly the chirality operator $\eta =C$
(see (\ref{2.49bb})--(\ref{2.51bb})).
By using the properties (\ref{2.5aaa}) of the Clifford algebra
and those of the chirality operator 
$\eta$ (see (\ref{2.49aaa})--(\ref{2.52aaa})), we compute
\be
\left[X+\lambda C\right]^2=2\lambda^2\II,
\label{2.95aac}
\ee
and
\be
\left[X+\lambda{1\over |u|}\Gamma(u)\right]^2=2\lambda\left[\lambda
-{g(\zeta,u)\over |u|}\right]\II .
\label{2.96aaa}
\ee
We see that, in both cases, the matrix $[X+\lambda(2P_D-\II)]$
is non-degenerate for any $(\zeta,\lambda)\ne (0,0)$ and 
any non-vanishing $\lambda$ that does not lie on the real axis.
 
Further, we find also
\be
\left[X+\lambda\Gamma(N)\right]^{2}
=2\lambda(\lambda -i\mu) \II .
\label{2.95aaa}
\ee
Bearing in mind that $\mu \equiv \sqrt{ |\zeta|^{2}
-\lambda^{2}}$, we see that this
does not vanish for any $\zeta\ne 0$. However, for $\zeta=0$ and 
{\it arbitrary} $\lambda$, with ${\rm Im}\,\lambda>0$, 
this equals zero. Thus, the boundary projector
$P_D=(\II\pm \Gamma(N))/2$ does not lead to strong ellipticity,
which only holds for ${\rm Im}\,\lambda<0$.

Thus, we have proven
\begin{theorem}
The boundary-value problem $(D,B_{D})$, where $D$ is a Dirac-type
operator and $B_{D}$ is a projector taking the forms 
(\ref{2.57aa}), or (\ref{2.56aa}), or (\ref{2.57aab}), 
is strongly elliptic with respect to
${\bf C}-\{0\}$, ${\bf C}-{\bf R}$ and the lower half-plane
${\rm Im}\,\lambda<0$, respectively.
\end{theorem}

\section{Resolvent and heat kernel}
\setcounter{equation}0

Let $w$ be a sufficiently large negative constant 
and $\lambda \in \CC$, ${\rm Re}\,\lambda<w$,
be a complex number with a sufficiently large negative real part.
Then the resolvent $G(\lambda)=(F-\lambda\II)^{-1}:\ L^2(V,M)\to L^2(V,M)$
of the strongly elliptic 
boundary-value problem $(F,B)$ is well defined. 
The {\it resolvent kernel} is a section of the tensor product 
of the vector bundles $V$ and $V^*$ over the tensor-product 
manifold $M\times M$, defined by the equation
\be
(F-\lambda\II)G(\lambda|x,y)=\delta(x,y)
\label{3.1nnn}
\ee
with the boundary conditions
\be
B_F\psi[G(\lambda|x,y)]=0,
\label{3.2nnn}
\ee
where $\delta(x,y)$ is the covariant Dirac distribution, 
which is nothing but the kernel 
of the identity operator.
Hereafter all differential operators as well 
as the boundary data map act on the {\it first}
argument of the resolvent kernel (and the heat kernel), 
unless otherwise stated.
This equation, together with the condition
\be
\overline{G(\lambda|x,y)}=G(\lambda^*|y,x),
\ee
which follows from the self-adjointness of the 
operator $F$, completely determine
the resolvent kernel.

Similarly, for $t>0$ the heat semi-group 
operator $U(t)=\exp(-tF):\ L^{2}(V,M)\to L^2(V,M)$
is well defined. The kernel of this operator, called heat kernel, 
is defined by the equation
\be
(\partial_t+F)U(t|x,y)=0
\ee
with the initial condition
\be
U(0|x,y)=\delta(x,y),
\ee
the boundary condition
\be
B_F\psi[U(t|x,y)]=0.
\ee
and the self-adjointness condition
\be
\overline{U(t|x,y)}=U(t|y,x).
\ee

As is well known \cite{gilkey95}, the heat 
kernel and the resolvent kernel are related
by the Laplace transform:
\be
G(\lambda|x,y)=\int\limits_0^\infty dt e^{t\lambda}U(t|x,y),
\label{3.8nnn}
\ee
\be
U(t|x,y)={1\over 2\pi i}\int\limits_{w-i\infty}^{w+i\infty}
d\lambda e^{-t\lambda}G(\lambda|x,y).
\ee

Also, it is well known \cite{gilkey95} that 
the heat kernel $U(t|x,y)$ is a smooth function near diagonal
of $M\times M$ and has a well defined diagonal value 
$U(t|x,x)$, and the functional trace
\be
\Tr_{L^2}\exp(-tF)=\int_M d\vol(x)\tr_V U(t|x,x).
\ee
Moreover, the functional trace has an asymptotic expansion  
as $t\to 0^+$
\be
\Tr_{L^2}\exp(-tF)\sim (4\pi t)^{-m/2}\sum\limits_{k\ge 0}
t^{k/2}A_{k/2}(F,B_F),
\ee
and the corresponding expansion of the 
functional trace for the resolvent as $\lambda\to-\infty$ reads
\be
\Tr_{L^2}\left({\partial\over \partial\lambda}\right)^n G(\lambda)
\sim (4\pi)^{-m/2}\sum\limits_{k\ge 0}\Gamma[(k-m)/2+n+1]
(-\lambda)^{(m-k)/2-n-1}A_{k/2}(F,B_F),
\ee
for $n\ge m/2$. 
Here $A_{k/2}(F,B_F)$ are the famous so-called 
({\it global}) heat-kernel coefficients
(sometimes called also Minakshisundaram-Plejel or Seeley coefficients).
The zeroth-order coefficient is very well known:
\be
A_0=\int\limits_M d\vol(x)\tr_V \II=\vol(M)\cdot\dim(V).
\label{3.13nnnz}
\ee
It is independent of the operator $F$ and of the boundary conditions.
The higher order coefficients have the following general form:
\be
A_{k/2}(F,B_F)=\int\limits_M d\vol(x)\tr_V a_{k/2}(F|x)
+\int\limits_{\partial M}d\vol(\hat x)\tr_V b_{k/2}(F,B_F|\hat x),
\ee
where $a_{k/2}$ and $b_{k/2}$ are the ({\it local}) 
{\it interior} and {\it boundary} 
heat-kernel coefficients. 
The interior coefficients do {\it not} 
depend on the boundary conditions $B_F$. 
It is well known that the interior coefficients of half-integer order, 
$a_{k+1/2}$, vanish \cite{gilkey95}.
The integer order coefficients $a_k$ are 
calculated for Laplace-type operators up to $a_4$ \cite{avra91b}.
The boundary coefficients $b_{k/2}(F,B_F)$ do 
depend on {\it both} the operator $F$
and the boundary operator $B_F$. 
They are far more complicated because in addition to the geometry
of the manifold $M$ they depend essentially on the geometry of the 
boundary $\partial M$. 
For Laplace-type operators they are known 
for the usual boundary conditions (Dirichlet,
Neumann, or mixed version of them) up to $b_{5/2}$ 
\cite{branson90,kirsten}. For 
generalized boundary conditions including tangential derivatives, 
to the best of our knowledge, 
they are not known at all. Only some 
special cases have been studied in the literature 
\cite{mcavity91,dowker97,avresp97}.

We are going to calculate below the next-to-leading coefficient 
$A_{1/2}(F,B_F)$ for the generalized
boundary conditions.
To do this, and also to study the role of the 
ellipticity condition, we will construct an approximation
to the heat kernel $U(t|x,y)$ near the
diagonal, i.e. for $x$ close to $y$ 
and for $t\to 0^{+}$. Since the heat kernel
and resolvent kernel are connected by the Laplace transform, this 
is equivalent to studying an approximation 
to the resolvent kernel $G(\lambda|x,y)$ near the diagonal and for 
large negative $\lambda\to -\infty$ (this leads, in turn, to an
approximate inverse of $F-\lambda \II$, called a {\it parametrix}). 

Let us stress here that we are not going to provide 
a rigorous construction of the resolvent with all 
the estimates, which, for the boundary-value problem, 
is a task that would require a separate paper. 
For a complete and mathematically rigorous 
exposition the reader is referred to the classical papers
\cite{gilkeysmith83b,seeley69a,seeley69b,gilkeysmith83a,
hormander85,booss93}.

Here we keep instead to a pragmatic approach and 
will describe briefly how the approximate resolvent kernel
for $\lambda\to -\infty$, and hence the heat kernel 
for $t\to 0^{+}$ can be constructed, and then will
calculate both kernels in the leading approximation. 
This will allow us to compute the heat-kernel coefficient $A_{1/2}$.

First of all, we decompose both kernels into two parts
\be
G(\lambda|x,y)=G_\infty(\lambda|x,y)+G_B(\lambda|x,y),
\ee
\be
U(t|x,y)=U_\infty(t|x,y)+U_B(t|x,y).
\ee
Then we construct {\it different } approximations for $G_\infty$
and $G_B$ and,
analogously, for $U_\infty$ and $U_B$.
The first parts $G_\infty(\lambda|x,y)$ and $U_\infty(t|x,y)$ are 
approximated by the usual asymptotic expansion of the 
resolvent and the heat kernel in the case of compact 
manifolds {\it without boundary} when
$x\to y$, $\lambda\to -\infty$ and $t\to 0^{+}$. 
This means that effectively one introduces a small expansion parameter 
$\varepsilon$ reflecting the fact that the points $x$ and $y$ 
are close to each other, the parameter
$t$ is small and the parameter $\lambda$ is negative and large. 
This can be done by fixing a point $x'$, choosing the normal 
coordinates at this point (with $g_{\mu\nu}(x')=\delta_{\mu\nu}$) and 
scaling
\be
x\to x'+\varepsilon(x-x'), \qquad y\to x'+\varepsilon(y-x'), 
\qquad t\to \varepsilon^2t,
\qquad \lambda\to \varepsilon^{-2}\lambda
\ee
and expanding into an asymptotic series in $\varepsilon$.
If one uses the Fourier transform, then the 
corresponding momenta $\xi\in T^*M$
are large and scale according to 
\be
\xi\to \varepsilon^{-1}\xi.
\ee
This construction is standard \cite{gilkey95} 
and we do not repeat it here. 
One can also use a completely covariant method 
\cite{avra91b,avra98}. Probably
the most convenient formula for the asymptotics as $t\to 0^{+}$, 
among many equivalent ones, is
\cite{avra91b,avra98}
\be
U_\infty(t|x,y)\sim (4\pi t)^{-m/2}\exp\left(-{\sigma\over 2t}\right)
\Delta^{1/2}
\sum\limits_{k\ge 0} {t^k}a_k,
\label{3.19xxx}
\ee
where $\sigma=\sigma(x,y)=r^2(x,y)/2$ is one half 
the geodesic distance between $x$ and $y$,
$\Delta=\Delta(x,y)=g^{-1/2}(x)g^{-1/2}(y)
\det(-\partial^x_\mu\partial^y_{\nu}\sigma(x,y))$ is
the corresponding Van Vleck-Morette determinant, $g=\det g_{\mu\nu}$, 
and $a_k=a_k(x,y)$ are the off-diagonal heat-kernel
coefficients. These coefficients satisfy certain 
differential recursion relations which can
be solved in form of a covariant Taylor series near diagonal \cite{avra91b}.
On the diagonal the asymptotic expansion 
of the heat kernel reads
\be
U_\infty(t|x,x)
\sim (4\pi t)^{-m/2}\sum\limits_{k\ge 0}t^{k}a_{k}(x,x).
\ee
The explicit formulas for the diagonal values of $a_k$ are known up 
to $k=4$ \cite{avra91b}. This asymptotic expansion can be integrated over
the manifold $M$ to get
\be
\int\limits_M d\vol(x)\tr_V U_\infty(t|x,x)
\sim (4\pi t)^{-m/2}\sum\limits_{k\ge 0}t^{k}
\int\limits_M d\vol(x)\tr_V\,a_{k}(x,x).
\ee
Thus, integrating the diagonal of $U_\infty$ gives the 
interior terms in the heat-kernel asymptotics. 

The asymptotic expansion of $G_\infty(\lambda|x,y)$ 
for $\lambda\to -\infty$ is obtained from here by 
the Laplace transform (\ref{3.8nnn})
\be
G_{\infty}(\lambda|x,y)\sim (4\pi)^{-m/2}\Delta^{1/2}
\sum\limits_{k\ge 0}2\left({\sigma\over -2\lambda}\right)^{(2k+2-m)/4}
K_{k+1-m/2}(\sqrt{-2\lambda\sigma}) a_k,
\label{3.20nnn}
\ee
where $K_\nu(z)=\pi/[2\sin(\nu\pi)][I_{-\nu}(z)-I_\nu(z)]$ 
is the McDonald function (the 
modified Bessel function of third kind). 
It is singular on the diagonal $x=y$.
However, for a sufficiently large $n$, $n\ge m/2$, it becomes
regular at the diagonal resulting in the 
asymptotic series as $\lambda\to -\infty$
\bea
&&\int\limits_M d\vol(x)\tr_V\left\{\left(
{\partial\over \partial\lambda}\right)^n 
G_\infty(\lambda|x,y)\right\}\Bigg|_{x=y}\nonumber\\
&&\sim (4\pi)^{-m/2}\sum\limits_{k\ge 0}\Gamma\left(k-{m\over 2}+n+1\right)
(-\lambda)^{{m\over 2}-k-n-1}
\int\limits_M d\vol(x)\tr_V\,a_{k}(x,x).
\eea

For a {\it strongly elliptic} boundary-value problem 
the diagonal of the boundary part $U_B(t|x,x)$ 
is {\it exponentially small}, i.e. of order $\sim\exp(-r^2(x)/t)$, 
where $r(x)$ is the normal geodesic
distance to the boundary, as $t\to 0^{+}$ if $x\not\in\partial M$.
So, it does not contribute to the asymptotic expansion of the heat-kernel
diagonal outside the boundary as $t \to 0^{+}$. 
Therefore, the asymptotic expansion of the total heat-kernel diagonal
outside the boundary is
determined only by $U_\infty$
\be
U(t|x,x)\sim (4\pi t)^{-m/2}\sum\limits_{k\ge 0} {t^k}a_k(x,x),
\qquad x\not\in \partial M.
\ee
The point is, the coefficients of the asymptotic expansion of
the diagonal of the boundary part $U_B(t|x,x)$ as $t\to 0^{+}$ behave near
the boundary like the one-dimensional Dirac 
distribution $\delta(r(x))$ and its derivatives.
Thus, the {\it integral} over the manifold $M$ of the boundary part
$U_B(t|x,x)$ has an asymptotic expansion as 
$t \to 0^{+}$ with non-vanishing coefficients in form of the integrals
over the boundary. Therefore, it determines the local boundary contributions 
$b_{k/2}$ to the heat-kernel coefficients $A_{k/2}$. 
It is well known that the coefficient $A_{1/2}$ is a purely boundary
contribution \cite{gilkey95}. It is almost obvious that it
can be evaluated by integrating the fibre trace of the boundary
contribution $U_{B}$ of the heat kernel in the leading order.

Of course, this approximation is obtained {\it without}
taking into account the boundary conditions. 
Therefore, $G_\infty$ satisfies approximately
the equation (\ref{3.1nnn}) but does not satisfy 
the boundary conditions (\ref{3.2nnn}).
This means that the compensating term $G_B(\lambda|x,y)$ 
is defined by the equation
\be
(F-\lambda\II)G_B(\lambda|x,y)=0
\label{3.13nnn}
\ee
with the boundary condition
\be
B_F\psi\left[G_\infty(\lambda|x,y)
+G_B(\lambda|x,y)\right]=0.
\label{3.14nnn}
\ee
Analogously, $U_B(t|x,y)$ is defined by
\be
(\partial_t+F)U_B(t|x,y)=0
\ee
with the initial condition
\be
U_B(0|x,y)=0,
\ee
and the boundary condition
\be
B_F\psi\left[U_\infty(t|x,y)
+U_B(t|x,y)\right]=0.
\ee

The most difficult problem is to find the compensating 
terms $G_B(\lambda|x,y)$ and $U_{B}(t|x,y)$.
These functions are important only near the boundary 
where they behave like distributions when
$t\to 0^{+}$ or $\lambda\to -\infty$. Since the points $x$
and $y$ are close to the boundary the coordinates $r(x)$ and $r(y)$ 
are small {\it separately}, hence not only
the difference $[r(x)-r(y)]$ but also the sum $[r(x)+r(y)]$ is small. 
This means that we must additionally scale $r(x)\to \varepsilon r(x)$, 
$r(y)\to \varepsilon r(y)$. By contrast,
the point $\hat x'$ is kept fixed on the boundary, so the
coordinates $\hat x'$ do not scale at all: $\hat x'\to \hat x'$.

Thus, we shall scale the coordinates 
$x=(\hat x,r(x))$ and $y=(\hat y,r(y))$, 
the  parameters $t$ and $\lambda$ and momenta $\xi=(\zeta,\omega)\in T^*M$, 
with $\zeta\in T^*\partial M$ and $\omega \in T^* {\bf R}$, according to
\be
\hat x\to \hat x'+\varepsilon(\hat x-\hat x'),\qquad 
\hat y\to \hat x'+\varepsilon(\hat y-\hat x'), \qquad 
r(x)\to \varepsilon r(x),\qquad 
r(y)\to \varepsilon r(y), 
\ee
\be
t\to \varepsilon^2 t, \qquad \lambda\to 
{1\over \varepsilon^2}\lambda,\qquad
\zeta \to {1\over \varepsilon}\zeta, \qquad 
\omega\to {1\over\varepsilon}\omega.
\ee
The corresponding differential operators are scaled by
\be
\hat\partial\to{1\over\varepsilon}\hat\partial, \qquad
\partial_r\to{1\over\varepsilon}\partial_r,\qquad
\partial_t\to{1\over\varepsilon^2}\partial_t,\qquad
\partial_\lambda\to\varepsilon^2\partial_\lambda.
\ee
We will call this transformation just {\it scaling} 
and denote the scaled objects by an index $\varepsilon$,
e.g. $G_B^{\varepsilon}$. The scaling parameter 
$\varepsilon$ will be considered as a small parameter
in the theory and we
will use it to expand everything in power series in $\varepsilon$.
We will {\it not} take care about the convergence 
properties of these expansions and take them as
{\it formal} power series. In fact, they are asymptotic 
expansions as $\varepsilon\to 0$.
At the very end of calculations we set $\varepsilon=1$. 
The non-scaled objects, i.e. those with
$\varepsilon=1$, will not have the index $\varepsilon$, 
e.g. $G_B^\varepsilon|_{\varepsilon=1}=G_B$.
Another way of doing this is by saying that we will 
expand all quantities in the homogeneous functions
of the coordinates $(\hat x-\hat x'), (\hat y-\hat x'), 
r(x), r(y)$, the momenta $\xi=(\zeta,\omega)$ and the parameters $t, \lambda$.

First of all, we expand the scaled operator $F^\varepsilon $ in power 
series in $\varepsilon$ 
\be
F^\varepsilon\sim \sum\limits_{n\ge 0}\varepsilon^{n-2} F_n,
\ee
where $F_n$ are second-order differential operators with homogeneous symbols.
The boundary operator requires a more careful handling. Since half of
the boundary data (\ref{6e}) contain
normal derivatives, formally $\psi_0=\varphi|_{r=0}$ 
and $\psi_1=\partial_r\varphi|_{r=0}$,
(\ref{7e}), would be of different order in $\varepsilon$. 
To make them of the same order we have to 
assume an additional factor $\varepsilon$ in all 
$\psi_1\in C^{\infty}(W_1,\partial M)$. Thus, we define 
the {\it graded scaling} of the boundary data map by
\be
\psi^\varepsilon(\varphi)=\left(\matrix{\psi^\varepsilon_0(\varphi)\cr
\varepsilon\psi^\varepsilon_1(\varphi)\cr}\right)
=\left(\matrix{\varphi(\hat x, r)|_{r=0}\cr
\partial_r\varphi(\hat x,r)|_{r=0}\cr}\right)=\psi(\varphi),
\label{25bnnn}
\ee
so that the boundary data map $\psi$ {\it does not scale} at all.
This leads to an additional factor $\varepsilon$ in 
the operator $\Lambda$ determining the
boundary operator $B_F$ (\ref{25e}). Thus, we 
define the {\it graded scaling} of the 
boundary operator by 
\be
B^\varepsilon_F=\left(\matrix{\Pi^\varepsilon&0\cr
\varepsilon\Lambda^\varepsilon& \II-\Pi^\varepsilon\cr}\right),
\label{25ennn}
\ee
which has the following asymptotic expansion in $\varepsilon$:
\be
B_F^\varepsilon\sim\sum\limits_{n\ge 0}\varepsilon^{n} B_{F(n)},
\ee
where $B_{F(n)}$ are first-order tangential operators 
with homogeneous symbols. Of course,
\be
F_0=-\partial_r^2-\hat\partial^2,
\label{3.35nnn}
\ee
\be
B_{F(0)}=\left(\matrix{\Pi_0&0\cr
\Lambda_0& \II-\Pi_0\cr}\right),
\label{25fnnn}
\ee
where 
\be
\hat\partial^2=\hat g^{jk}(\hat x')\hat\partial_{j}
\hat\partial_{k},\qquad
\Lambda_0=\Gamma^j(\hat x')\hat \partial_j,\qquad
\Pi_0=\Pi(\hat x').
\label{3.37nnn}
\ee
Note that all leading-order operators $F_0$, $B_{F(0)}$ 
and $\Lambda_0$ have {\it constant}
coefficients and, therefore, are very easy to handle. 
This procedure is called sometimes 
``freezing the coefficients of the differential operator''.

The subsequent strategy is rather simple. 
Expand the scaled resolvent kernel in $\varepsilon$
\be
G^\varepsilon_\infty\sim\sum_{n\ge 0}
\varepsilon^{2-m+n}G_{\infty(n)},
\ee
\be
G^\varepsilon_B\sim\sum_{n\ge 0}\varepsilon^{2-m+n}G_{B(n)},
\ee
and substitute into the scaled version of the equation (\ref{3.13nnn})  
and the boundary condition (\ref{3.14nnn}). Then, by equating  the like 
powers in $\varepsilon$ one gets an infinite set of recursive
equations which determine all $G_{B(n)}$. 
The zeroth-order term $G_{B(0)}$ is defined by
\be
(F_0-\lambda\II)G_{B(0)}=0,
\ee
and the boundary conditions, 
\be
B_{F(0)}\psi\left(G_{\infty(0)}+G_{B(0)}\right)=0,
\ee
where $F_0$ and $B_{F(0)}$ are defined by (\ref{3.35nnn})-(\ref{3.37nnn}).
The higher orders are determined from
\be
(F_0-\lambda \II)G_{B(k)}=-\sum\limits_{n=1}^{k}F_nG_{B(k-n)}, \qquad
k=1,2,\dots,
\ee
and the corresponding boundary conditions, 
\be
B_{F(0)}\psi\left(G_{\infty(k)}+G_{B(k)}\right)
=-\sum\limits_{n=1}^{k}B_{F(n)}\psi\left(G_{\infty(k-n)}
+G_{B(k-n)}\right).
\ee
The $G_{\infty(n)}$ are obtained simply by expanding 
the scaled version of (\ref{3.20nnn}) in power series in $\varepsilon$.
The operator $F_0$ is a partial differential operator, 
but fortunately, has constant coefficients. By
using the Fourier transform in the boundary coordinates 
$(\hat x-\hat x')$ it reduces to an {\it ordinary}
differential operator of second order. This enables one 
to find easily its resolvent. We will do this in next
subsection below. Before doing this, let us stress that the 
same procedure can be applied to get the
boundary part $U_{B}(t|x,y)$ of the heat kernel. 

\subsection{Leading order resolvent}

In this subsection we determine the resolvent to leading order, 
i.e. $G_0=G_{\infty(0)}+G_{B(0)}$.
As we already outlined above, we fix a point 
$\hat x'\in \partial M$ on the boundary and the
normal coordinates at this point (with $\hat 
g_{ik}(\hat x')=\delta_{ik}$), take
the tangent space $T\partial M $ and identify the manifold $M$ with 
$M_0\equiv T\partial M\times{\bf R}_{+}$.
By using the explicit form of the zeroth-order operators 
$F_0$, $B_0$ and $\Lambda_0$
given by (\ref{3.35nnn})-(\ref{3.37nnn})
we obtain the equation
\be
\left(-\partial_{r}^{2}-\hat\partial^2-\lambda\right)G_0(\lambda|x,y)
=\delta(x-y),
\label{3.44nnn}
\ee
and the boundary conditions
\be
\Pi_0G_0(\lambda|x,y)\Big|_{r(x)=0}=0,
\label{3.45nnn}
\ee
\be
(\II-\Pi_0)\left(\partial_r+i\Gamma^j_0\hat\partial_j\right)
G_0(\lambda|x,y)\Big|_{r(x)=0}=0,
\label{3.46nnn}
\ee
where $\Pi_0=\Pi(\hat x'), \Gamma^j_0=\Gamma^j(\hat x')$. 
Hereafter the differential operators always act 
on the first argument of a kernel. Moreover,
for simplicity of notation, we will denote $\Pi_0$ and $\Gamma_0$ 
just by $\Pi$ and $\Gamma^j$ and omit the dependence 
of all geometric objects
on $\hat x'$. To leading order this cannot cause any 
misunderstanding.
Further to this, the resolvent kernel should be bounded,
\be
\lim_{r(x)\to\infty}G_0(\lambda|x,y)
=\lim_{r(y)\to\infty}G_0(\lambda|x,y)=0 ,
\ee
and self-adjoint,
\be
\overline{G_0(\lambda|x,y)}
={G_0(\lambda^*|y,x)}.
\ee
Since the above boundary-value problem has constant coefficients,
by using the Fourier transform in $(\hat x-\hat y)$
\be
G_{(0)}(\lambda|x,y)=\int\limits_{{\bf R}^{m-1}}{d\zeta\over (2\pi)^{m-1}}
e^{i\zeta\cdot(\hat x-\hat y)}\tilde G_{(0)}(\lambda|\zeta,r(x),r(y)),
\ee
where $\zeta\in T^*\partial M$, 
$\zeta\cdot\hat x\equiv \zeta_j\hat x^j$, we obtain
an ordinary differential equation
\be
\left(-\partial_r^2+|\zeta|^2-\lambda\right)\tilde G_0(\lambda|\zeta,r(x),r(y))
=\delta[\rho(x,y)],
\label{3.44nnnm}
\ee
where $|\zeta|^2=g^{ij}(\hat x')\zeta_i\zeta_j$, 
\be
\rho(x,y)=r(x)-r(y),
\ee
the boundary conditions
\be
\Pi\tilde G_0(\lambda|\zeta,0,r(y))=0,
\label{3.45nnnm}
\ee
\be
(\II-\Pi)\left(\partial_r+iT\right)
\tilde G_0(\lambda|\zeta,r(x),r(y))\Big|_{r(x)=0}=0,
\label{3.46nnnm}
\ee
and self-adjointness condition
\be
\overline{G_0(\lambda|\zeta,r(x),r(y))}
={G_0(\lambda^*|\zeta,r(y),r(x))}.
\ee
Here $T$ is an anti-self-adjoint matrix
\be
T=T(\zeta)=\Gamma\cdot\zeta=\Gamma^j\zeta_j, \qquad \bar T=-T,
\label{3.55nna}
\ee
satisfying the conditions
\be
(\II-\Pi)T=T(\II-\Pi)=T.
\label{3.55nnb}
\ee

The first `free' part of the solution of this problem $\tilde G_{\infty(0)}$, 
which gives the resolvent kernel on the line ${\bf R}$,  is almost obvious.
Since the equation (\ref{3.44nnnm}) has constant 
coefficients and there are no boundary conditions, 
its solution depends only on the difference $\rho(x,y)=r(x)-r(y)$ 
\bea
\tilde G_{\infty(0)}(\lambda|\zeta,r(x),r(y))
&=&{1\over 2\mu}\left\{\theta[\rho(x,y)]e^{-\mu\rho(x,y)}
+\theta[-\rho(x,y)]e^{\mu\rho(x,y)}\right\}
\nonumber\\
&=&{1\over 2\mu}e^{-\mu|\rho(x,y)|}.
\label{3.55nnn}
\eea
where
\be
\mu \equiv \sqrt{|\zeta|^2-\lambda},\qquad {\rm Re}\,\mu>0,
\ee
and $\theta$ is the usual step function
\be
\theta(x) \equiv \left\{
\begin{array}{ll}
1 & {\rm if}\ x>0\\
0 & {\rm if} \ x<0. 
\end{array}\right.
\ee
The boundary part ${\tilde G}_{B(0)}$ takes into account the
boundary conditions (\ref{3.45nnnm}) and (\ref{3.46nnnm}). 
On requiring regularity at infinity
and self-adjointness of the kernel, 
we can write down the solution up to a 
constant {\it self-adjoint} matrix $h=\bar h$,
\be
\tilde G_{B(0)}(\lambda|\zeta,r(x),r(y))
={1\over 2\mu}h(\mu;\zeta)\,e^{-\mu[r(x)+r(y)]}.
\label{3.58nnn}
\ee
>From eqs. (\ref{3.55nnn})-(\ref{3.58nnn}) we find
\be
\tilde G_0(\lambda|\zeta,0,r(y))={1\over 2\mu}(\II+h)e^{-\mu r(y)},\qquad
\partial_r\tilde G_0(\lambda|\zeta,r(x),r(y))\Big|_{r(x)=0}
={1\over 2}(\II-h)e^{-\mu r(y)}.
\ee
Taking into account the conditions 
(\ref{3.55nna})-(\ref{3.55nnb}) we reduce
the boundary conditions to the form
\be
\Pi(\II+h)=0,
\label{3.62nna}
\ee
\be
(\II-\Pi)\mu(\II-h)+iT(\II+h)=0.
\label{3.63nna}
\ee
Since $h$ is self-adjoint and $T\Pi=\Pi T=0$ 
it must have the form $h=\alpha\Pi+(\II-\Pi)\beta(T)(\II-\Pi)$,
where $\alpha$ is a constant and $\beta$ is a matrix depending on $T$.
Substituting this into (\ref{3.62nna}) we immediately obtain 
$\alpha=-1$. From the second boundary condition (\ref{3.63nna}) we get 
\be
(\mu\II-iT)\beta=\mu\II+iT.
\ee
Since the problem is assumed to be strongly elliptic, the matrix 
$\mu\II-iT$ is invertible for any $(\zeta,\lambda)\ne (0,0)$.
Thus, we obtain the solution of the boundary conditions in the form
\bea
h(\mu;\zeta)&=&-\Pi+(\II-\Pi)(\mu\II-iT)^{-1}(\mu\II+iT)(\II-\Pi)
\nonumber\\
&=&\II -2\Pi
+2iT(\mu\II-iT)^{-1}.
\label{572k}
\eea
In a particular case of mixed Dirichlet and Neumann boundary 
conditions, when $\Gamma^{j}=0$, 
this reduces to 
\be
h=\II-2\Pi.
\ee
Note that the function $h(\mu,\zeta)$ depends 
actually on the ratio $\zeta/\mu$,
in other words, it satisfies the homogeneity relation
\be
h(s\mu,s\zeta)=h(\mu,\zeta).
\label{142a}
\ee
More generally,
\be
\tilde G_{0}\left({1\over t}\lambda\Big|
{1\over \sqrt t}\zeta,{\sqrt t}\,r(x),{\sqrt t}\,r(y)\right)
=\sqrt t\tilde G_{0}(\lambda|\zeta,r(x),r(y))
\ee
This holds for $G_{\infty(0)}$ as well as for $G_{B(0)}$.
As a function of $\lambda$, $\tilde G_{(0)}$ is a meromorphic
function with a cut along the positive real axis from $|\zeta|^2$ to $\infty$.
The ``free'' part $G_{\infty(0)}$ has no other singularities, whereas the
boundary part $G_{B(0)}$ has simple poles at the eigenvalues of the
matrix $|\zeta|^2+T^2$. Remember that the strong ellipticity condition
\ref{52eaa} requires these eigenvalues to be 
positive and smaller than $|\zeta|^2$,
i.e. $0<|\zeta|^2+T^2<|\zeta|^2$.

\subsection{Leading order heat kernel}

Using $G_0(\lambda|x,y)$, 
we can also get the zeroth-order heat kernel $U_0(t|x,y)$
by the inverse Laplace transform
\bea
U_0(t|x,y)&=&{1\over 2\pi i}\int\limits_{w-i\infty}^{w+i\infty} 
d\lambda\, e^{-t\lambda}G_0(\lambda|x,y)\\
&=&\int\limits_{{\bf R}^{m-1}} 
{d\zeta\over (2\pi)^{m-1}}
\int\limits_{w-i\infty}^{w+i\infty} 
{d\lambda\over 2\pi i}\,
e^{-t\lambda +i\zeta\cdot(\hat x-\hat y)}
\tilde G_0(\lambda|\zeta,r(x),r(y))
\eea
where $w$ is a negative constant.
Now, by scaling the integration variables $\lambda\to \lambda/t$
and $\zeta\to \zeta/\sqrt t$ and shifting the contour of integration
over $\lambda$ ($w\to w/t$, which can be done because the integrand
is analytic in the left half-plane of $\lambda$) and
using the homogeneity property we obtain immediately
\be
U_0(t|x,y)=(4\pi t)^{-m/2}\int\limits_{{\bf R}^{m-1}} 
{d\zeta\over \pi^{(m-1)/2}}\,e^{i\zeta\cdot{(\hat x-\hat y)/\sqrt t}}
\int\limits_{w-i\infty}^{w+i\infty} 
{d\lambda\over  i\sqrt \pi}\, e^{-\lambda}\,
\tilde G_0\left(\lambda\Big|\zeta,{r(x)\over\sqrt t},
{r(y)\over \sqrt t}\right)
\ee
Next, let us change the variable $\lambda$ according to
\be
\lambda \equiv |\zeta|^2+\omega^2,
\ee 
where $\omega=i\mu$, and, hence, ${\rm Im}\,\omega>0$. In the upper
half-plane, ${\rm Im}\,\omega>0$, this change of variables is single valued
and well defined. Under this change the cut in the complex plane $\lambda$
along the positive real axis from $|\zeta|^2$ to $\infty$, i.e. 
${\rm Im}\,\lambda=0,\quad|\zeta|^2<{\rm Re}\, 
\lambda<\infty$ is mapped onto the
whole real axis ${\rm Im}\,\omega=0,\quad -\infty<{\rm Re}\,\omega<+\infty$. 
The interval ${\rm Im}\,\lambda=0,\quad 0<{\rm Re}\, \lambda<|\zeta|^2$ 
on the real axis of $\lambda$ is mapped onto an
interval ${\rm Re}\,\omega=0,\quad 0<{\rm Im}\,\omega<|\zeta|$,
on the {\it positive imaginary} axis of $\omega$. 
As a function of $\omega$ the resolvent
$\tilde G_0$ is a meromorphic function in the upper half plane, 
${\rm Im}\,\omega>0$, with simple poles on the interval 
${\rm Re}\,\omega=0,\quad 0<{\rm Im}\,\omega<|\zeta|$, on the
imaginary axis. The contour of integration in the complex plane of $\omega$
is a hyperbola going from $(e^{i3\pi/4})\infty$ through the point 
$\omega=\sqrt{|\zeta|^2-w}$ to
$(e^{i\pi/4})\infty$. It   
can be deformed to a contour $C$ that comes from 
$-\infty+i\varepsilon$, encircles the point 
$\omega=i|\zeta|$ in the clockwise direction 
and goes to $+\infty+i\varepsilon$, where $\varepsilon$ is an
infinitesimal positive parameter. The contour 
$C$ does {\it not} cross the interval
${\rm Re}\,\omega=0,\quad 0<{\rm Im}\,\omega<|\zeta|$, on the
imaginary axis and is {\it above} all the singularities of the resolvent.

After such a tranformation we obtain
\bea
U_0(t|x,y)&=&(4\pi t)^{-m/2} 
\int\limits_{{\bf R}^{m-1}} 
{d\zeta\over \pi^{(m-1)/2}}e^{-|\zeta|^2
+i\zeta\cdot{(\hat x-\hat y)/\sqrt t}}
\nonumber\\
&\times&
\int\limits_{C} 
{d\omega\over\sqrt\pi}\, e^{-\omega^2}\,2(-i\omega)\,
\tilde G_0\left(|\zeta|^2+\omega^2\Big|\zeta,
{r(x)\over\sqrt t},{r(y)\over \sqrt t}\right)
\eea
Substituting here $\tilde G_0$ we find
\bea
U_0(t|x,y)&=&(4\pi t)^{-m/2} 
\int\limits_{{\bf R}^{m-1}} 
{d\zeta\over \pi^{(m-1)/2}}e^{-|\zeta|^2
+i\zeta\cdot{(\hat x-\hat y)/\sqrt t}}
\nonumber\\
&\times&
\int\limits_{C} 
{d\omega\over\sqrt\pi}\, e^{-\omega^2}\,
\left\{e^{i\omega|r(x)-r(y)|/\sqrt t}
+h(-i\omega;\zeta)e^{i\omega|r(x)+r(y)| / \sqrt t}\right\}.
\eea
The first ``free'' part is easily obtained by computing the Gaussian 
integrals over $\omega$ and $\zeta$. We get
\be
U_0(t|x,y)=(4\pi t)^{-m/2}\exp\left(-{1\over 4t}|x-y|^2\right)
+U_{B(0)}(t|x,y),
\ee
where
\bea
U_{B(0)}(t|x,y)&=&(4\pi t)^{-m/2} 
\int\limits_{{\bf R}^{m-1}} 
{d\zeta\over \pi^{(m-1)/2}}e^{-|\zeta|^2
+i\zeta\cdot{(\hat x-\hat y)/\sqrt t}}
\\
&\times&
\int\limits_{C} 
{d\omega\over\sqrt\pi}\, 
\exp\left\{-\omega^2+i\omega{|r(x)+r(y)|\over\sqrt t}\right\}\,
\left\{(\II-2\Pi)-2T(\omega\,\II+T)^{-1}\right\}.
\nonumber
\eea
Here, the part proportional to $(\II-2\Pi)$ also contains only 
Gaussian integrals
and is easily calculated. Thus (cf. \cite{avra93})
\be
U_{B(0)}(t|x,y)=(4\pi t)^{-m/2}\left\{
\exp\left[-{1\over 4t}|\hat x-\hat y|^2
-{1\over 4t}[r(x)+r(y)]^2\right] (\II-2\Pi)
+\Omega(t|x,y)\right\},
\label{152}
\ee
where 
\bea
\Omega(t|x,y)&=&
-2\int\limits_{{\bf R}^{m-1}} {d\zeta\,\over \pi^{(m-1)/2}}\,
\int\limits_C{d\omega\over\sqrt\pi} 
\exp\left\{-|\zeta|^2+i\zeta\cdot{(\hat x-\hat y)\over\sqrt t}
-\omega^2+i\omega{[r(x)+r(y)]\over\sqrt t}\right\}
\nonumber\\[12pt]
&&\times
\Gamma\cdot\zeta
(\omega\,\II+\Gamma\cdot\zeta)^{-1},
\eea
where we substituted $T=\Gamma\cdot\zeta\equiv\Gamma^j\zeta_j$.

Herefrom we obtain easily the diagonal value of the heat kernel 
\be
U_{(0)}(t|x,x)=(4\pi t)^{-m/2}\left\{\II+
\exp\left(-{r^2(x)\over t}\right) (\II-2\Pi)
+\Phi(r(x)/\sqrt t)\right\},
\label{152pp}
\ee
where 
\bea
\Phi(z)&=&
-2\int\limits_{{\bf R}^{m-1}} {d\zeta\,\over \pi^{(m-1)/2}}\,
\int\limits_C{d\omega\over\sqrt\pi}\, 
e^{-|\zeta|^2-\omega^2+2i\omega z}\,
\Gamma\cdot\zeta
(\omega\,\II+\Gamma\cdot\zeta)^{-1}.
\eea
Now, by using the representation (remember that 
${\rm Im}\,(\omega\,\II+\Gamma\cdot\zeta)>0$)
\be
(\omega\II +\Gamma\cdot\zeta)^{-1}=
-2i\int_0^\infty dp\,e^{2ip(\omega\II +\Gamma\cdot\zeta)},
\label{383}
\ee
and computing a Gaussian integral over $\omega$, we obtain
\be
\Phi(z)=
2\int\limits_{{\bf R}^{m-1}} {d\zeta \,\over \pi^{(m-1)/2}}\,
\int_0^\infty dp\,
e^{-|\zeta|^2-(p+z)^2}
{\partial\over\partial p}
e^{2ip\Gamma\cdot\zeta}.
\ee

Integrating by parts over $p$ we get
\be
\Phi(z)=-2e^{-z^2}\II-2{\partial\over\partial z}\Psi(z),
\label{3.85nnnz}
\ee
where
\be
\Psi(z)=\int\limits_{{\bf R}^{m-1}} {d\zeta \,\over \pi^{(m-1)/2}}\,
\int_0^\infty dp\,
e^{-|\zeta|^2-(p+z)^2+2ip\Gamma\cdot\zeta}.
\label{3.86nnny}
\ee
It is not difficult to show that, as $z\to \infty$, 
the functions $\Psi(z)$ and $\Phi(z)$ are {\it exponentially
small}:
\be
\Psi(z)\sim {1\over 2z}e^{-z^2} \left[\II
-{1\over 2z^{2}}(\II+\Gamma^{2})+{\rm O}(z^{-4})\right],
\ee
\be
\Phi(z)\sim {1\over z^2}e^{-z^2} \left[-\Gamma^{2}
+{\rm O}(z^{-2})\right],
\label{3.87aaaz}
\ee
where $\Gamma^{2} \equiv g_{ij}\Gamma^{i}\Gamma^{j}$.
For $z=0$, by using the change $\zeta\to -\zeta$, we obtain
\be
\Psi(0)={\sqrt\pi\over 2}\int\limits_{{\bf R}^{m-1}}
{d\zeta \,\over \pi^{(m-1)/2}}\,
e^{-|\zeta|^2-(\Gamma\cdot\zeta)^2}.
\label{3.87nnnz}
\ee
Note that 
this integral converges when the strong ellipticity condition
$(\Gamma\cdot\zeta)^2+|\zeta|^2>0$ is satisfied.

Now, we take the diagonal
$U_{(0)}(t|x,x)$ given by (\ref{152pp}), 
and integrate over the manifold $M$. 
Because the boundary part $U_{B(0)}$ is exponentially small as
$r(x)\to \infty$ we can in fact integrate it only over a narrow
strip near the boundary, when $0<r(x)<\delta$. The difference
is asymptotically small as $t\to 0^{+}$. Doing the change of variables
$z=r/\sqrt t$ we reduce the integration to $0<z<\delta/\sqrt t$.
We see, that as $t\to 0^{+}$ we can integrate over $z$ from $0$ to $\infty$.
The error is asymptotically small as $t\to 0^{+}$ and does not contribute to
the asymptotic expansion of the trace of the heat kernel.

Thus, we obtain
\be
\Tr_{L^2} U_{0}(t)\sim(4\pi t)^{-m/2}\left(A_0+\sqrt t\, A_{1/2}
+\cdots\right),
\ee
where $A_0$ is given by (\ref{3.13nnnz}) and
\be
A_{1/2}=\int\limits_{\partial M}d\vol(\hat x)\tr_V a_{1/2},
\ee
with
\be
a_{1/2}=
{\sqrt\pi\over 2}\left(\II-2\Pi+\beta_{1/2}
\right),
\label{104f}
\ee
\be
\beta_{1/2}={2\over\sqrt\pi}\int\limits_0^\infty
dz\,\Phi(z).
\ee
Now, using (\ref{3.85nnnz}) and (\ref{3.87nnnz})
and the fact that $\Psi(\infty)=0$
we get easily
\bea
\beta_{1/2}&=&-2+{4\over \sqrt \pi}\Psi(0)
\nonumber\\
&=&
2\int\limits_{{\bf R}^{m-1}} {d\zeta \,\over \pi^{(m-1)/2}}\,
e^{-|\zeta|^2}
\left(e^{-A^{jk}\zeta_j\zeta_k}-\II\right),
\label{165f}
\eea
where 
\be
A^{jk} \equiv \Gamma^{(j}\Gamma^{k)}.
\ee
Eventually,
\be
a_{1/2}={\sqrt\pi\over 2}\left\{\II-2\Pi+
2\,\int\limits_{{\bf R}^{m-1}} {d\zeta \,\over \pi^{(m-1)/2}}\,
e^{-|\zeta|^2}
\left(e^{-A^{jk}\zeta_j\zeta_k}-\II\right)\right\}.
\label{3.45ee}
\ee

Further calculations of general nature, without knowing the algebraic
properties of the matrices $A^{jk}$, seem to be impossible.
One can, however, evaluate the integral in form of an expansion
in the matrices $A^{jk}$, or $\Gamma^i$.
Using the Gaussian integrals
\be
\int\limits_{{\bf R}^{m-1}} {d\zeta \,\over \pi^{(m-1)/2}}\,
e^{-|\zeta|^2}\zeta_{i_1}\cdots\zeta_{i_{2n}}
={(2n)!\over n!2^{2n}}\hat g_{(i_{1}i_{2}}
\cdots\hat g_{i_{2n-1}i_{2n})},
\ee
we obtain
\be
{\beta_{1/2}}=
2\sum_{n\ge 1}
{(-1)^n}{(2n)!\over (n!)^22^{2n}}
\hat g_{i_1i_2}\cdots\hat g_{i_{2n-1}i_{2n}}
\Gamma^{(i_1}\cdots \Gamma^{i_{2n})},
\ee
and, therefore,
\be
a_{1/2}={\sqrt\pi\over 2}\Biggl\{\II-2\Pi
+2\sum_{n\ge 1}
{(-1)^n}{(2n)!\over (n!)^22^{2n}}
\hat g_{i_1i_2}\cdots\hat g_{i_{2n-1}i_{2n}}
\Gamma^{(i_1}\cdots \Gamma^{i_{2n})}\Biggr\}.
\label{3.49ee}
\ee

Since our main result (\ref{165f}) is rather complicated, we now consider
a number of particular cases of physical relevance.

\begin{enumerate}

\item
The simplest case is, of course, when 
the matrices $A^{ij}$ vanish, $A^{ij}=0$. 
One then gets the familiar
result for mixed boundary conditions 
\cite{gilkey95,branson90}
\be
\beta_{1/2}=0,\qquad 
a_{1/2}={\sqrt\pi\over 2} (\II-2\Pi).
\ee

\item
The first non-trivial case is when the 
matrices $\Gamma^i$ form an Abelian algebra,
\be
[\Gamma^i,\Gamma^j]=0.
\ee
One can then easily compute the integral (\ref{165f}) explicitly,
\bea
\beta_{1/2}=2\left[(\II+\Gamma^2)^{-1/2}-1\right].
\eea
Therefore
\be
a_{1/2}={\sqrt\pi\over 2} \left\{\II-2\Pi
+2\left[(\II+\Gamma^2)^{-1/2}
-\II\right]\right\}.
\label{167}
\ee

In the case $\Pi=0$, 
this coincides with the result of ref. \cite{mcavity91},
where the authors considered the particular case of commuting
$\Gamma^{i}$ matrices (without noting this explicitly).
 
\item
Let us assume that the matrices $A^{jk}=\Gamma^{(j}\Gamma^{k)}$ 
form an Abelian algebra, i.e.
\be
[A^{jk},A^{lm}]=0.
\label{168}
\ee

The evaluation of the resulting integral over $\zeta$ yields
\bea
\beta_{1/2}=2\left[
(\det_{T\partial M } [\delta^{i}{}_j+A^{i}{}_j])^{-1/2}-\II \right],
\eea
where the determinant $\det_{T\partial M }$ is taken over the indices
in the tangent space to the boundary (we used $\det\hat g=1$).
By virtue of (\ref{104f}) one gets the final result
\be
a_{1/2}={\sqrt\pi\over 2} \left\{\II-2\Pi
+2\left[\left(\det_{T\partial M } [\delta^{i}{}_j+A^{i}{}_j]\right)^{-1/2}
-\II\right]\right\}.
\label{167f}
\ee

\item
A particular realization of the last case is
when the matrices $\Gamma^i$ satisfy the Dirac-type condition
\be
\Gamma^{i}\Gamma^{j}+\Gamma^{j}\Gamma^{i}
=2\,\hat g^{ij}{1\over (m-1)}\Gamma^{2} ,
\ee
so that
\be
A^{ij}=\Gamma^{(i}\Gamma^{j)}
={1\over (m-1)}\hat g^{ij}\Gamma^2.
\ee
Then, of course, the matrices $A^{ij}$ commute and the result is given
by eq. (\ref{167f}). But the determinant is now
\be
\det_{T\partial M } [\delta^{i}{}_j+A^{i}{}_j]
=\left(\II+{1\over (m-1)}\Gamma^2\right)^{m-1}.
\ee
Thus,
\be
a_{1/2}={\sqrt\pi\over 2} \left\{\II-2\Pi
+2\left[\left(\II+{1\over (m-1)}\Gamma^2\right)^{-(m-1)/2}
-\II\right]\right\}.
\label{167c}
\ee
Note that this {\it differs essentially} from the 
result of ref. \cite{mcavity91}, and shows again that the
result of ref. \cite{mcavity91} applies actually only to the completely
Abelian case, when all matrices $\Gamma^j$ commute.
Note also that, in the most interesting applications 
(e.g. in quantum gravity), the
matrices $\Gamma^{i}$ do not commute \cite{avresp97}.
The result (\ref{165f}), however, is valid in the most general case.

\item
A very important case is when the operator 
$\Lambda$ is a natural operator on the boundary.
Since it is of first order it can be only 
the generalized Dirac operator. In this
case the matrices $\Gamma^j$ satisfy the condition
\be
\Gamma^{i}\Gamma^{j}+\Gamma^{j}\Gamma^{i}\equiv 2A^{ij}
=-2\,\kappa\hat g^{ij}(\II-\Pi),
\label{3.60eee}
\ee
where $\kappa$ is a constant.
Hence the matrices $A^{ij}$ obviously commute 
and we have the case considered above
(see (\ref{167f})).
The determinant is easily calculated,
\be
\det_{T\partial M }[\delta^i_j+A^i_j]=\Pi+(\II-\Pi)(1-\kappa)^{m-1},
\ee
and we eventually obtain 
\be
a_{1/2}={\sqrt\pi\over 2} \left\{-\Pi
+(\II-\Pi)\left[2(1-\kappa)^{-(m-1)/2}-1\right]\right\}.
\label{167e}
\ee
Thus, a singularity is found at $\kappa=1$. 
This happens because, for $\kappa=1$, the strong ellipticity
condition is violated (see also \cite{dowker97}).
Indeed, the strong ellipticity condition (\ref{52eaaa}), 
\be
T^{2}+|\zeta|^2\II=(\Gamma\cdot\zeta)^2+|\zeta|^2\II
=|\zeta|^2[\Pi+(1-\kappa)(\II-\Pi)]>0,
\ee
implies in this case $\kappa<1$ (cf. \cite{dowker97}).

\end{enumerate}

\section{Analysis of ellipticity in a 
general gauge theory on manifolds with boundary}
\setcounter{equation}0

In this section we are going to study gauge-invariant 
boundary conditions in a general gauge theory
(for a review, see \cite{dewitt83}).
A gauge theory is defined by two vector bundles, 
$V$ and $G$, such that $\dim V>\dim G$.
$V$ is the bundle of gauge fields $\varphi\in C^\infty(V,M)$, 
and $G$ is the bundle 
of parameters of gauge transformations $\epsilon\in C^\infty(G,M)$.
Both bundles $V$ and $G$ are equipped with some Hermitian 
positive-definite metrics $E$, $E^{\dag}=E$, and $\gamma$, 
$\gamma^{\dag}=\gamma$,
and with the corresponding natural $L^2$ scalar products
$(,)_V$ and $(,)_G$.

The infinitesimal gauge transformations
\be
\delta \varphi=R\epsilon
\label{202}
\ee
are determined by a first-order differential operator $R$,
\be
R:\ C^\infty(G,M)\to C^\infty(V,M).
\ee
Further, one introduces two auxiliary operators,
\be
{X}:\ C^\infty(V,M)\to C^\infty(G,M)
\label{1997}
\ee
and
\be
{Y}:\ C^\infty(G,M)\to C^\infty(G,M),
\ee
and one defines two differential operators,
\be
{L} \equiv {X} R:\ C^\infty(G,M)\to C^\infty(G,M) 
\label{211}
\ee
and
\be
H \equiv \bar{X}{Y}{X}:\ C^\infty(V,M)\to C^\infty(V,M),
\label{209}
\ee
where $\bar{X}=E^{-1}{X}^{\dag}\gamma$.
The operators $X$ and $Y$ should satisfy the following conditions
(but are otherwise arbitrary):
\begin{itemize}
\item[1)]
The differential operators $L$ and $H$ have the same order.
\item[2)]
The operators $L$ and $H$ are formally self-adjoint (or anti-self-adjoint).
\item[3)]
The operators $L$ and $Y$ are elliptic.

\end{itemize}

{}From these conditions we find that there are 
two essentially different cases:

\paragraph{Case I.}
$X$ is of first order and $Y$ is of zeroth order, i.e.
\be
X=\bar R, \qquad Y=\II_G,
\ee
where $\bar R \equiv \gamma^{-1}R^{\dag}E$. Then, of course,
$L$ and $H$ are both {\it second-order} differential operators,
\be
L=\bar R R, \qquad H=R\bar R.
\ee

\paragraph{Case II.}
${X}$ is of zeroth order and ${Y}$ is of first order.
Let ${\cal R}$ be the bundle of maps of $G$ into $V$, and let
$\beta\in {\cal R}$ be a zeroth-order differential operator. Then
\be
X=\bar \beta, \qquad Y=\bar\beta R ,
\ee
where $\bar\beta \equiv \gamma^{-1}\beta^{\dag}E$,
and the operators $L$ and $H$ are of {\it first} order,
\be
L=\bar\beta R, \qquad H=\beta\bar\beta R\bar\beta=\beta L\bar\beta.
\ee
We assume that, by suitable choice of the parameters,
the second-order operator $\bar R R$ can be made
of Laplace type and the first-order operator 
$\bar\beta R$ can be made of Dirac type,
and, therefore, have non-degenerate leading symbols, 
\be
\det_G\sigma_L(\bar R R)\ne 0,
\ee
\be
\det_G\sigma_L(\bar \beta R)\ne 0.
\ee
The dynamics of gauge fields $\varphi\in C^\infty(V,M)$ 
at the linearized (one-loop) level is described by a
formally self-adjoint (or anti-self-adjoint) differential operator,
\be
\Delta:\ C^\infty(V,M)\to C^\infty(V,M).
\ee
This operator is of second order for bosonic fields and of 
first order for fermionic fields.
In both cases it satisfies the identities
\be
\Delta R = 0,\qquad \bar R\Delta=0,
\label{221}
\ee
and, therefore, is degenerate.

We consider only the case when the gauge generators are
{\it linearly independent}. This means that the equation
\be
\sigma_L(R)\epsilon=0,
\ee
for $\xi\ne 0$, has {\it the only} solution $\epsilon=0$. 
In other words, 
\be
{\rm Ker\,}\sigma_L(R)=\emptyset,
\ee
i.e. the rank
of the leading symbol of the operator $R$ equals the dimension
of the bundle $G$,
\be
\rank \sigma_L(R)=\dim G.
\label{215g}
\ee
We also assume that the leading symbols of the generators $R$ 
are {\it complete} in that they generate {\it all} 
zero-modes of the leading
symbol of the operator $\Delta$, i.e. {\it all} solutions of the equation
\be
\sigma_L(\Delta)\varphi=0,
\ee
for $\xi\ne 0$, have the form
\be
\varphi=\sigma_L(R)\epsilon,
\ee
for some $\epsilon$. In other words,
\be
{\rm Ker\,}\sigma_L(\Delta)
=\{\sigma_L(R)\epsilon\ |\ \epsilon\in G\},
\ee
and hence
\be
\rank\sigma_L(\Delta)=\dim V-\dim G.
\ee
Further, let us take the operator $H$ of the 
{\it same} order as the operator $\Delta$
and construct a formally (anti-)self-adjoint operator, 
\be
F \equiv \Delta+H,
\label{1230h}
\ee
so that
\be
\sigma_L(F)=\sigma_{L}(\Delta)+\sigma_{L}(H).
\ee

It is easy to derive the following result: 
\begin{proposition}
The leading symbol of the operator $F$ is non-degenerate, i.e.
\be
\det_V\sigma_L(F;x,\xi)\ne 0,
\ee
for any $\xi\ne 0$.
\end{proposition}

\paragraph{Proof.}
Indeed, suppose there exists a zero-mode, 
say $\varphi_0$, of the leading symbol
of the operator $F$, i.e.
\be
\sigma_L(F)\varphi_0=\bar\varphi_0\sigma_L(F)=0,
\ee
where
$\bar\varphi\equiv \varphi^{\dag}E$.
Then we have
\be
\bar\varphi_0\sigma_L(F)\sigma_L(R)
=\bar\varphi_0\sigma_L(\bar XY)\sigma_L({L})=0,
\ee
and, since $\sigma_L({L})$ is non-degenerate,
\be
\bar\varphi_0\sigma_L(\bar XY)=\sigma_L(YX)\varphi_0=0.
\label{235}
\ee
But this implies
\be
\sigma_L(H)\varphi_0=0,
\ee
and hence
\be
\sigma_L(F)\varphi_0=\sigma_L(\Delta)\varphi_0=0.
\ee
Thus, $\varphi_0$ is a zero-mode of the leading 
symbol of the operator $\Delta$,
and according to the completeness of the generators $R$ must have the form
$\varphi_0=\sigma_L(R)\epsilon$ for some $\epsilon$.
Substituting this form into the eq. (\ref{235}) we obtain
\be
\sigma_L(YX)\sigma_L(R)\epsilon
=\sigma_L({YL})\epsilon=0.
\label{235a}
\ee
Herefrom, by taking into account the non-singularity of $\sigma_L({YL})$,
it follows $\epsilon=\varphi_0=0$, and hence the leading symbol of the 
operator $F$ does not have any zero-modes, i.e. it is non-degenerate.

Thus, the operators ${L}$ and $F$ have, 
both, non-degenerate leading symbols.
In quantum field theory the operator $X$ is called the gauge-fixing operator, 
$F$ the gauge-field operator, the operator ${L}$ 
the (Faddeev-Popov) ghost operator
and the operator $Y$ in the Case II the third 
(or Nielsen-Kallosh) ghost operator.
The most convenient and the most important case is when,
by suitable choice of the parameters it turns out to be possible to
make both the operators $F$ and ${L}$ either of Laplace type
or of Dirac type. The one-loop effective action for  
gauge fields is given by the functional
superdeterminants of the gauge-field operator $F$ 
and the ghost operators $L$ and $Y$ \cite{dewitt83}
\be
\Gamma={1\over 2}\log({\rm Sdet}\, F)
-\log({\rm Sdet}\,L)
-{1\over 2}\log({\rm Sdet}\,Y).
\ee

\subsection{Bosonic gauge fields}

Let us consider first the case of bosonic fields, when $\Delta$ is a 
second-order formally self-adjoint operator. 
The gauge invariance identity (\ref{221}) means, in particular,
\be
\sigma_L(\Delta)\sigma_L(R) = 0.
\label{227}
\ee
Now we assume that both the operators $L=\bar RR$ and $F=\Delta+R\bar R$ 
are of Laplace type, i.e.
\be
\sigma_L(\bar R R)=|\xi|^{2}\II_{G},
\label{4.25aa}
\ee
\be
\sigma_L(F)=\sigma_L(\Delta)+\sigma_L(R\bar R)=|\xi|^{2}\II .
\label{245g}
\ee
On manifolds with boundary one has to impose some boundary conditions 
to make these operators self-adjoint and elliptic. They read
\be
B_L\psi(\epsilon)=0,
\label{246h}
\ee
\be
B_F\psi(\varphi)=0,
\label{880h}
\ee
where $\psi(\epsilon)$ and $\psi(\varphi)$ are the 
boundary data for the bundles
$G$ and $V$, respectively, and $B_{L}$ and $B_F$ are the corresponding
boundary operators (see section 2.1). 
In gauge theories one tries to choose the 
boundary operators $B_{L}$ and $B_F$
in a gauge-invariant way, so that the condition
\be
B_F\psi(R\epsilon) = 0
\ee
is satisfied identically 
for any $\epsilon$ subject to the boundary conditions (\ref{246h}).
This means that the boundary operators $B_L$ and $B_F$ satisfy the identity
\be
B_F [\psi,R](\II-B_{L})\equiv 0,
\ee
where $[\psi,R]$ is the commutator of the linear 
boundary data map $\psi$ and the operator $R$.

We will see that this requirement fixes completely the form of the
as yet unknown boundary operator $B_{L}$. Indeed,
the most natural way to satisfy the condition 
of gauge invariance is as follows.
Let us decompose the cotangent bundle 
$T^*M $ in 
such a way that $\xi=(N,\zeta)\in T^*M $, where 
$N$ is the inward pointing unit normal to the boundary and
$\zeta\in T^*\partial M $ is a cotangent vector on the boundary.
Consider the restriction $W_0$ of the vector bundle $V$ to the boundary.
Let us define restrictions of the leading symbols 
of the operators $R$ and $\Delta$ to the boundary, i.e.
\be
\Pi\equiv \sigma_L(\Delta;N)\Big|_{\partial M} ,
\ee
\be
\nu\equiv \sigma_L(R;N)\Big|_{\partial M},
\label{234x}
\ee
\be
\mu\equiv \sigma_L(R;\zeta)\Big|_{\partial M}.
\label{235x}
\ee
{}From eq. (\ref{227}) we have thus the identity
\be
\Pi\nu=0,
\label{179j}
\ee
Moreover, from (\ref{4.25aa}) and (\ref{245g}) we have also
\be
\bar\nu\nu=\II_G ,
\label{264g}
\ee
\be
\bar\nu\mu+\bar\mu\nu=0,
\label{268g}
\ee
\be
\bar\mu\mu=|\zeta|^2\II_G ,
\label{269g}
\ee
\be
\Pi=\II -\nu\bar\nu .
\label{261g}
\ee
{}From (\ref{179j}) and (\ref{264g})
we find that $\Pi:\ W_0\to W_0$ is a self-adjoint 
projector orthogonal to $\nu$,
\be
\Pi^2=\Pi,\qquad \Pi\nu=0, \qquad \bar\Pi=\Pi.
\ee
Then, a part of the boundary conditions for the operator $F$ reads
\be
\Pi\varphi\Big|_{\partial M}=0.
\label{257}
\ee
The gauge transformation of this equation is
\be
\Pi R\epsilon\Big|_{\partial M}=0.
\label{257ee}
\ee
The normal derivative does not contribute to
this equation, and, therefore, if 
Dirichlet boundary conditions are imposed on $\epsilon$,
\be
\epsilon\Big|_{\partial M}=0,
\label{221eee}
\ee
the equation (\ref{257ee}) is satisfied identically.

The easiest way to get the other part of the 
boundary conditions is just to set
\be
\bar R\varphi\Big|_{\partial M}=0.
\label{221g}
\ee
Bearing in mind eq. (\ref{211}) we find that, 
under the gauge transformations (\ref{202}), 
this is transformed into
\be
{L}\epsilon\Big|_{\partial M}=0.
\ee
If some $\epsilon$ is a zero-mode of the operator 
${L}$, i.e. $\epsilon\in {\rm Ker}\,({L})$, 
this is identically zero. For all $\epsilon\notin   {\rm Ker}\,({L})$ 
this is identically zero for the Dirichlet boundary conditions
(\ref{221eee}). In other words, the requirement of gauge
invariance of the boundary conditions 
(\ref{880h}) determines in an almost unique way (up to zero-modes)
that the ghost boundary operator $B_L$ should be of Dirichlet type. 
Anyway, the Dirichlet boundary conditions
for the operator $L$ are {\it sufficient} to achieve
gauge invariance of the boundary conditions for the operator $F$.

Since the operator ${\bar R}$ in the boundary conditions 
(\ref{221g}) is a first-order
operator, the set of boundary conditions (\ref{257}) and (\ref{221g})
is equivalent to the general scheme formulated in section 2.1.
Separating the normal derivative in the operator 
${\bar R}$ we find exactly the generalized boundary
conditions (\ref{14e}) with the boundary operator 
$B_F$ of the form (\ref{25e}) with a first-order operator
$\Lambda: C^\infty(W_0,\partial M)\to C^\infty(W_0,\partial M)$, 
the matrices $\Gamma^j$ being of the form
\be
\Gamma^j=-\nu\bar\nu\mu^j\bar\nu.
\label{278}
\ee
These matrices are anti-self-adjoint, $\bar\Gamma^i=-\Gamma^i$, 
and satisfy the relations
\be
\Pi\Gamma^i=\Gamma^i\Pi=0.
\ee
Thus, one can now define the matrix 
\be
T \equiv \Gamma\cdot\zeta 
=-\nu\bar\nu\mu\bar\nu,
\label{1460h}
\ee
where $\mu \equiv \mu^j\zeta_j$, and 
study the condition of strong ellipticity (\ref{52eaa}).
The condition of strong ellipticity now reads
\be
|\zeta|\II-iT=|\zeta|\II+i\nu\bar\nu\mu\bar\nu>0.
\ee

Further, using the eqs. (\ref{278}), 
(\ref{268g}) and (\ref{269g}) we evaluate
\be
A^{ij}=\Gamma^{(i}\Gamma^{j)}
=-(\II-\Pi)\mu^{(i}\bar\mu^{j)}(\II-\Pi).
\ee
Therefore,
\be
{T}^2=A^{ij}\zeta_i\zeta_j=
-(\II-\Pi)\mu\bar\mu(\II-\Pi),
\label{199j}
\ee
and
\be
T^2+|\zeta|^2\II=|\zeta|^2\Pi+(\II-\Pi)[|\zeta|^2\II
-\mu\bar\mu](\II-\Pi).
\ee
Since for non-vanishing $\zeta$ the part proportional 
to $\Pi$ is positive-definite,
the condition of strong ellipticity for bosonic gauge theory means
\be
(\II-\Pi)[|\zeta|^2\II-\mu\bar\mu](\II-\Pi)>0.
\ee

We have thus proved a theorem:

\begin{theorem}
Let $V$ and $G$ be two vector bundles over a compact Riemannian 
manifold $M$ with smooth boundary, such that
$\dim V>\dim G$. Consider a bosonic gauge theory and let the 
first-order differential operator
$R:\ C^\infty(G,M)\to C^\infty(V,M)$ be the generator of infinitesimal
gauge transformations.
Let $\Delta:\ C^\infty(V,M)\to C^\infty(V,M)$ be the 
gauge-invariant second-order differential operator of the linearized
field equations. Let the
operators ${L} \equiv \bar R R:\ C^\infty(G,M)\to C^\infty(G,M)$
and $F \equiv \Delta+R\bar R$ be of Laplace type and normalized by 
$\sigma_L({L})=|\xi|^2\II_G$.
Let $\sigma_L(R;N)\Big|_{\partial M}\equiv \nu$
and $\sigma_L(R;\zeta)\Big|_{\partial M}\equiv \mu$ be the restriction
of the leading symbol of the operator $R$ to the boundary,
$N$ being the normal to the boundary
and $\zeta\in T^{*}\partial M$ being a cotangent vector, and 
$\Pi=\II -\nu\bar\nu$.

Then the generalized boundary-value problem $(F,B_F)$ with the boundary
operator $B_F$ determined by the boundary 
conditions (\ref{257}) and (\ref{221g})
is gauge-invariant provided that the ghost boundary operator 
$B_{L}$ takes the Dirichlet form. 
Moreover, it is strongly elliptic with respect to ${\bf C}
-{\bf R}_{+}$
if and only if the matrix $[|\zeta|\II+i\nu\bar\nu\mu\bar\nu]$
is positive-definite.
A sufficient condition for that reads
\be
(\II-\Pi)[|\zeta|^2\II-\mu\bar\mu](\II-\Pi)>0.
\label{203j}
\ee
\end{theorem}

\subsection{Fermionic gauge fields}

In the case of fermionic gauge fields, $\Delta$ is a first-order formally
self-adjoint (or anti-self-adjoint), degenerate (i.e. gauge-invariant) 
operator with leading symbol satisfying
\be
\sigma_{L}(\Delta)\sigma_L(R)=0.
\label{4.79aaa}
\ee
Now we have the case II, hence, we assume both the operators $L=\bar\beta R$
and $F=\Delta+\beta\bar\beta R\bar\beta$ to be of Dirac type, i.e.
the operators $L^2$ and $F^2$ are of Laplace type:
\be
[\bar\beta\sigma_L(R)]^2=|\xi|^2\II_{G},
\label{4.62}
\ee
\be
\left[\sigma_L(\Delta)+\beta\bar\beta 
\sigma_L(R)\bar\beta\right]^2=|\xi|^2\II .
\label{4.63}
\ee
Note that now we have {\it two} systems of Dirac matrices, $\gamma^\mu$
on the bundle $G$ and $\Gamma^\mu$ on $V$. 
They are defined by the leading symbols
of the Dirac-type operators $L$ and $F$, 
\be
\sigma_L(L)=-\gamma^\mu\xi_\mu, \qquad 
\sigma_L(F)=-\Gamma^\mu\xi_{\mu}.
\ee
Let us define
\be
b \equiv \sigma_L(\Delta;N)\Big|_{\partial M},\qquad
c \equiv \sigma_L(\Delta;\zeta)\Big|_{\partial M},
\ee
where $\zeta\in T^*\partial M $. Bearing in mind the notation
(\ref{234x}) and (\ref{235x}), we have from (\ref{4.79aaa})
\be
b\nu=0, \qquad c\nu+b\mu=0.
\label{4.84aa}
\ee
Moreover, from (\ref{4.62}) we have also
\be
\bar\beta\nu\bar\beta\nu=\II_G,
\ee
\be
\bar\beta\nu\bar\beta\mu+\bar\beta\mu\bar\beta\nu=0.
\ee
Similarly, from (\ref{4.63}) we get
\be
\left(b+\beta\bar\beta \nu\bar\beta\right)^{2}=\II ,
\ee
\be
\left(b+\beta\bar\beta \nu\bar\beta\right)
\left(c+\beta\bar\beta \mu\bar\beta\right)
+\left(c+\beta\bar\beta \mu\bar\beta\right)
\left(b+\beta\bar\beta \nu\bar\beta\right)=0.
\ee
Herefrom, we find that the following operators:
\be
P_G={1\over 2}[\II_{G}-\gamma(N)]={1\over 2}(\II_G+\bar\beta\nu),
\ee
\be
P_V={1\over 2}[\II -\Gamma(N)]
={1\over 2}(\II +b+\beta\bar\beta\nu\bar\beta),
\ee
are projectors,
\be
P_{G}^{2}=P_{G}, \qquad P_{V}^{2}=P_{V}.
\ee

The boundary conditions for the Dirac-type operators $L$ and $F$ 
are given by projectors (see (\ref{2.32aa})),
\be
P_{L}\epsilon\Big|_{\partial M}=0,
\label{4.85aaa}
\ee
\be
P_{F}\varphi\Big|_{\partial M}=0.
\label{4.85aa}
\ee
The problem is to make them gauge-invariant. Here the projectors
$P_{L}$ and $P_{F}$ are defined by means of the matrices 
$\gamma^{\mu}$ and $\Gamma^{\mu}$, respectively.
The gauge transformation of the equation (\ref{4.85aa}) is
\be
P_FR\epsilon\Big|_{\partial M}=0.
\label{4.86aa}
\ee
Noting that, on the boundary, 
\be
R\epsilon\Big|_{\partial M}=
[\nu\bar\beta\nu L
-i(\II -\nu\bar\beta\nu\bar\beta)\mu^j\hat\nabla_j]
\epsilon\Big|_{\partial M},
\ee
and assuming that $\mu$ commutes with $P_{L}$, we get
herefrom two conditions on the boundary projectors,
\be
P_F\nu\bar\beta\nu(\II_G-P_L)=0,
\ee
\be
P_F(\II -\nu\bar\beta\nu\bar\beta)(\II_G-P_L)=0.
\ee
Such gauge-invariant boundary operators always exist. We will construct them
explicitly in section 6 in the course of 
studying the Rarita--Schwinger system.

\section{Strong ellipticity in Yang--Mills theory}
\setcounter{equation}0
The first physical application that we study is the strong
ellipticity condition in Yang--Mills theory.
Now $G=A$ is the Lie algebra of a semi-simple and compact gauge group, and
$V$ is the bundle of $1$-forms taking values in $A$, i.e. $V=T^*M \otimes G$.
Let $h$ be the Cartan metric on the Lie algebra defined by
\be
h_{ab} \equiv -C^c{}_{ad}C^d{}_{cb},
\ee
$C^a{}_{bc}$ being the structure constants of the gauge group,
and the fibre metric $E$ on the bundle $V$ be defined by
\be
E(\varphi,\varphi) \equiv -\tr_A g(\varphi,\varphi),
\ee
or, in components,
\be
E^\mu{}_a{}^\nu{}_b=g^{\mu\nu}h_{ab}.
\ee
The Cartan metric is non-degenerate and 
positive-definite. Therefore, the 
fibre metric $E$ is always non-singular and positive-definite.
Henceforth we will suppress the group indices. The generator
of gauge transformations is now just the covariant derivative
\be
R=\nabla^G,
\qquad
\bar R=-\tr_g\nabla^V.
\ee
The leading symbols of these operators are
\be
\sigma_L(R;\xi)=i\xi\II_G, \qquad
\sigma_L(\bar R;\xi)=-i\bar\xi\II  ,
\ee
where
$\bar\xi \equiv \tr_g\xi$
is a map $\bar\xi:\ T^*M \to {{\bf R}}$.
First of all, we see that
\be
\sigma_L(L;\xi)=\sigma_L(\bar R R;\xi)
=\bar\xi\xi \II_G=|\xi|^2\II_{G},
\ee
\be
\sigma_L(H;\xi)=\sigma_L(R\bar R;\xi)
=\xi\otimes\bar\xi\II ,
\ee
so that the operator $L=\bar R R$ is indeed a Laplace-type operator.
The gauge-invariant operator $\Delta$ in linearized Yang--Mills theory
is defined by the leading symbol
\be
\sigma_L(\Delta)=\left(|\xi|^2-\xi\otimes
\bar\xi\right)\II .
\ee
Thus, the operator $F=\Delta+H$ is of Laplace type,
\be
\sigma_L(F)=\sigma_L(\Delta+H)=|\xi|^2\II .
\ee
Further, we find
\be
\nu=i\,N\II_G, \qquad \bar\nu=-i\,\tr_g N ,
\label{210j}
\ee
and
\be
\mu=i\,\zeta\II_G, \qquad \bar\mu=-i\,\tr_g\bar\zeta ,
\label{211j}
\ee
where $N$ is the normal cotangent vector, and $\zeta\in T^*\partial M $.

The projector $\Pi$ has the form
\be
\Pi=q,
\ee
where
\be
q\equiv 1-N\otimes\bar N .
\label{5.13cc}
\ee
In components, this reads
\be
q_\mu{}^\nu\equiv \delta^\nu_\mu-N_\mu N^\nu. 
\ee
Thus, the gauge-invariant boundary conditions are
\be
q_\mu{}^\nu \varphi_\nu\Big|_{\partial M}=0,
\label{197e}
\ee
\be
g^{\mu\nu}\nabla_\mu\varphi_\nu\Big|_{\partial M}=0.
\label{198e}
\ee
Since $\bar\zeta N=\bar N\zeta=<\zeta,N>=0$, 
we find from (\ref{210j}) and (\ref{211j})
\be
\bar\mu\nu=\bar\nu\mu=0,
\ee
and hence the matrices $\Gamma^i$ in (\ref{55}),
as well as ${T}=\Gamma^i\zeta_i$, vanish:
\be
\Gamma^j=0, \; \; T=0.
\ee
Therefore, the matrix ${T}^2+ \zeta^{2} \II
=|\zeta|^2\II$ is positive-definite,
so that the strong ellipticity condition 
(\ref{52eaa}) or (\ref{203j}) is satisfied.

Thus, we have
\begin{theorem}
A Laplace-type operator $F$ acting on $1$-forms taking values in a 
semi-simple Lie algebra $G$,
$F: \ C^\infty(T^{*}M\otimes G,M)\to C^\infty(T^{*}M\otimes G,M)$,
with the boundary conditions (\ref{197e}) and (\ref{198e}), 
is elliptic.
\end{theorem}

Since such boundary conditions automatically appear in the 
gauge-invariant formulation of the boundary conditions in one-loop
Yang--Mills theory, we have herefrom

\begin{corollary}
The boundary-value problem in one-loop Euclidean Yang--Mills theory
determined by a Laplace-type operator and the 
gauge-invariant boundary conditions defined by 
(\ref{197e}) and (\ref{198e}), is strongly elliptic with respect
to ${\bf C}-{\bf R}_{+}$.
\end{corollary}

\section{Ellipticity for the Rarita--Schwinger system}
\setcounter{equation}0

The next step is the analysis of Rarita--Schwinger fields.
The bundle $G$ is now the bundle of spinor fields taking values in some 
semi-simple Lie algebra $A$, i.e. $G={\cal S}\otimes A$, and $V$ is
the bundle of spin-vector fields, in other words, $V$ is 
the bundle of $1$-forms taking values in the fibre of $G$,
i.e. $V=T^*M \otimes G$. The whole theory does not depend on the presence
of the algebra $A$, so we will omit completely the group indices.

Let $\gamma^\mu$ be the Dirac matrices 
and $\gamma$ be the Hermitian 
metric on the spinor bundle ${\cal S}$ determined by
\be
\gamma^\mu\gamma^\nu+\gamma^\nu\gamma^\mu=2g^{\mu\nu}\II_{S},
\qquad
\bar\gamma^{\mu}=-\varepsilon\gamma^{\mu},
\qquad
\gamma^{\dag}=\gamma,
\label{225k}
\ee
where $\bar\gamma^\mu=\gamma^{-1}\gamma^\mu{}^{\dag}\gamma$.
The fibre metric $E$ on the bundle $V$ is defined by
\be
E(\varphi,\varphi) \equiv \varphi^{\dag}_\mu 
\gamma E^{\mu\nu}\varphi_{\nu} ,
\ee
where
\be
E^{\mu\nu}\equiv g^{\mu\nu}+\alpha\gamma^{\mu}\gamma^{\nu},
\ee
with a parameter $\alpha$. By using
(\ref{225k}) it is easily seen that $E$ is Hermitian, i.e.
\be
\bar E^{\mu\nu}=E^{\nu\mu},
\ee
if $\alpha$ is real.
The inverse metric reads
\be
E^{-1}_{\mu\nu}=g_{\mu\nu}
-{\alpha\over (1+m\alpha)}\gamma_{\mu}\gamma_{\nu}.
\ee
Therefore, the fibre metric $E$ is positive-definite
only for $\alpha>-1/m$ and is singular for $\alpha=-1/m$.
Thus, hereafter $\alpha\ne -1/m$.
In fact, for $\alpha=-1/m$ the matrix $E_\mu{}^\nu$ becomes a projector
on a subspace of spin-vectors $\varphi_\mu$ satisfying the condition
$\gamma^\mu\varphi_\mu=0$.

The generator of gauge transformations is now 
again, as in the Yang--Mills case,
just the covariant derivative \cite{avra91}
\be
R=b\nabla^G, 
\ee
where $b$ is a normalization constant, with leading symbol
\be
\sigma_L(R;\xi)=ib\xi\II_{G}.
\ee
Now we have the Case II of section 4, and hence 
we define the map $\beta:\ G\to V$ and
its adjoint $\bar\beta:\ V\to G$ by
\be
(\beta\epsilon)_{\mu} \equiv
{i\varepsilon\over b}E^{-1}_{\mu\nu}\gamma^\nu\epsilon
={i\varepsilon\over b(1+\alpha m)}\gamma_\mu\epsilon, 
\qquad
\bar\beta\varphi \equiv {i\over b}\gamma^\mu\varphi_\mu,
\ee
so that
\be
\bar\beta\beta={-\varepsilon m\over b^{2} (1+\alpha m)}\II_{G}.
\ee

The operator $X$ of Eq. (\ref{1997}) is 
now equal to $\bar\beta$ so that
the operator $L=\bar\beta R$ is of Dirac type with leading symbol
\be
\sigma_L(L)=-\gamma^\mu\xi_\mu.
\ee
The leading symbol of the operator $H=\beta\bar\beta R\bar\beta$ reads
\be
\sigma_L(H)\varphi_{{\lambda}}
={\varepsilon\over b^2(1+\alpha m)}
\gamma_{{\lambda}}\gamma^\mu\gamma^\nu\xi_\mu\varphi_{\nu}.
\label{236j}
\ee
The gauge-invariant operator $\Delta$ is now
the Rarita--Schwinger operator with leading symbol
\be
\sigma_L(\Delta)\varphi_{{\lambda}}=\varepsilon E^{-1}_{{\lambda}\beta}
\gamma^{\mu\beta\nu}\xi_\mu\varphi_{\nu}.
\ee
Here and below we denote the antisymmetrized products 
of $\gamma$-matrices by
\be
\gamma^{\mu_1\dots\mu_n}\equiv \gamma^{[\mu_1}\cdots\gamma^{\mu_n]}.
\ee
Of course, the leading symbol is self-adjoint 
and gauge-invariant, in that
\be
\overline{\sigma_L(\Delta)}=\sigma_L(\Delta),\qquad 
\sigma_L(\Delta)\sigma_L(R)=0,
\ee
where $\overline{\sigma_L(\Delta)}=E^{-1}\sigma^{\dag}_L(\Delta)E$.
Further, the leading symbol of the operator $F=\Delta+H$ reads
\be
\sigma_L(F)\varphi_{{\lambda}}
=\varepsilon\left[E^{-1}_{{\lambda}\beta}\gamma^{\mu\beta\nu}
+{1\over b^2(1+\alpha m)}\gamma_{{\lambda}}\gamma^{\mu}\gamma^{\nu}
\right]\xi_{\mu}\varphi_{\nu}.
\ee

Using the properties of the Clifford algebra we compute
\be
\sigma_L(F)\varphi_{{\lambda}}
=-\varepsilon \left\{
\delta^{\nu}_{{\lambda}}\gamma^\mu
-\delta^{\mu}_{{\lambda}}\gamma^\nu
+{1\over (1+m\alpha)}\left(1+2\alpha-{1\over b^2}\right)
\gamma_{{\lambda}}\gamma^\mu\gamma^\nu
-{(1+2\alpha)\over (1+m\alpha)}\gamma_{{\lambda}} g^{\mu\nu}
\right\}\xi_\mu\varphi_\nu.
\label{6.16aa}
\ee
Moreover, one can prove the following property of representation theory:
\begin{proposition}
The matrices 
$\widetilde{\Gamma}^\mu(0)$ defined by
\be
(\widetilde{\Gamma}^\mu(0))_{\lambda}{}^{\nu} = 
\widetilde{\Gamma}^\mu{}_{\lambda}{}^\nu(0)
\equiv \delta^\nu_{\lambda}\gamma^\mu
+{2\over m}\gamma_{\lambda}\gamma^\mu\gamma^\nu
+(\omega-1)\delta^\mu_{\lambda}\gamma^\nu
-(\omega+1)\gamma_{{\lambda}} g^{\mu\nu},
\label{6.20}
\ee
where $\omega$ is defined by
\be
\omega^{2} \equiv {m-4\over m},
\label{6.21ggg}
\ee
form a representation of the Clifford algebra, i.e.
\be
{\widetilde \Gamma}^\mu{}_\beta{}^{{\lambda}}(0)
{\widetilde \Gamma}^\nu{}_{\lambda}{}^{\rho}(0)
+{\widetilde \Gamma}^\nu{}_\beta{}^{{\lambda}}(0)
{\widetilde \Gamma}^\mu{}_{\lambda}{}^{\rho}(0)
=2g^{\mu\nu}\delta_{\beta}^{\rho}\II_{\cal S}.
\label{6.22ggg}
\ee
\end{proposition}

It is thus clear that the set of matrices
\be
\widetilde{\Gamma}^\mu{}_{\lambda}{}^\beta(\alpha')
=T^{-1}{}_{\lambda}{}^\rho(\alpha')
\widetilde{\Gamma}^\mu{}_\rho{}^\sigma(0)
T_\sigma{}^\beta(\alpha'),
\ee
with arbitrary non-degenerate matrix $T(\alpha')$ depending on
a parameter $\alpha'$, also forms a representation of the Clifford
algebra. By choosing
\be
T_{\lambda}{}^\beta(\alpha')=\delta_{{\lambda}}^\beta
+\alpha'\gamma_{{\lambda}}\gamma^\beta,
\ee
and, hence,
\be
T^{-1}{}_{\lambda}{}^\beta(\alpha')=\delta_{{\lambda}}^\beta
-{\alpha'\over (1+m\alpha')}\gamma_{{\lambda}}\gamma^\beta,
\ee
we prove, more generally, what follows.

\begin{corollary}
The matrices 
$\widetilde{\Gamma}^\mu(\alpha')$ defined by
\be
(\widetilde{\Gamma}^\mu(\alpha'))_{\lambda}{}^{\nu} = 
\widetilde{\Gamma}^\mu{}_{\lambda}{}^\nu(\alpha')
\equiv \delta^\nu_{\lambda}\gamma^\mu
+c_1\gamma_{\lambda}\gamma^\mu\gamma^\nu
+c_2\delta^\mu_{\lambda}\gamma^\nu
+c_3\gamma_{{\lambda}} g^{\mu\nu},
\label{6.20ggg}
\ee
where
\be
c_{1} \equiv {(m-2-m\omega)\over (1+m\alpha')}\left(\alpha'
+{\omega+1\over 2}\right)^2,
\ee
\be
c_{2} \equiv -(m-2-m\omega)\left(\alpha'+{\omega+1\over 2}\right),
\ee
\be
c_{3} \equiv -{2\over (1+m\alpha')}\left(\alpha'
+{\omega+1\over 2}\right),
\ee
with $\alpha'$ an arbitrary constant $\alpha'\ne -1/m$, 
and $\omega$ defined by (\ref{6.21ggg}),
form a representation of the Clifford algebra, 
i.e. satisfy the eq. (\ref{6.22ggg}).

\end{corollary}
Note that, by choosing $\alpha'=-(\omega+1)/2$, all the constants $c_1$, 
$c_2$ and $c_3$ vanish.

The operator $F$ with leading symbol (\ref{6.16aa}) should be of
Dirac type. Thus, by imposing that the term in curly brackets in Eq.
(\ref{6.16aa}) should coincide with the right-hand side of 
Eq. (\ref{6.20ggg}), one finds a system whose solution is
\be
\alpha'={1\over 4}\left[m-4+(m-2)\omega\right],
\ee
\be
\alpha={m-4\over 4},
\ee
\be
b=\pm {2\over \sqrt{m-2}}.
\ee
The simplest case is $m=4$, one has then 
$\alpha=\alpha'=0$, $b=\pm \sqrt 2$.

After this choice the operator $F$ is of Dirac type, i.e.
$
\sigma_L(F)=-\varepsilon\Gamma^\mu\xi_\mu,
$
with
\be
(\Gamma^\mu)_{\lambda}{}^\nu = 
\Gamma^\mu{}_{\lambda}{}^\nu
\equiv \delta^\nu_{\lambda}\gamma^\mu
+{1\over (m-2)}\gamma_{\lambda}\gamma^\mu\gamma^\nu
-\delta^\mu_{\lambda}\gamma^\nu
-{2\over (m-2)}\gamma_{{\lambda}} g^{\mu\nu}.
\label{6.20hhh}
\ee
Thus, we have two Dirac-type operators, $L$ and $F$,
which have elliptic leading symbols.
By choosing the appropriate boundary conditions with the projectors
$P_L$ and $P_F$ the system becomes elliptic. The problem
is to define the boundary projectors in a gauge-invariant way.

Let $P_L$ the boundary projector for the ghost operator $L$,
\be
P_{L}\epsilon\Big|_{\partial M}=0.
\label{6.28a}
\ee
Remember that it satisfies the symmetry condition (\ref{2.42aa}). 
Then we choose the boundary conditions for the gauge field
in the form
\be
P_L q_\nu{}^\mu\varphi_\mu\Big|_{\partial M}=0,
\label{6.28}
\ee
\be
P_L\gamma^\mu\varphi_\mu\Big|_{\partial M}=0,
\label{6.29}
\ee
where $q$ is a projector defined by (\ref{5.13cc}). 
The gauge transformation of (\ref{6.28}) reads
\be
P_Lq_\nu{}^\mu\nabla_\mu\epsilon\Big|_{\partial M}=0,
\ee
and does not include the normal derivative. We are assuming 
that the projector $P_{L}$ commutes with the tangential
derivative (as is usually the case, we find that their commutator
vanishes identically by virtue of the boundary conditions on $\epsilon$).
The gauge transformation of eq. (\ref{6.29}) is proportional 
exactly to the operator $L$,
\be
P_{L}\bar\beta R\epsilon\Big|_{\partial M}
=P_{L} L\epsilon\Big|_{\partial M}=0.
\ee
By expanding $\epsilon$ in the eigenmodes of the operator $L$ we find
that this is proportional to the boundary 
conditions (\ref{6.28a}) on $\epsilon$, and
therefore vanishes.
Thus, the boundary conditions (\ref{6.28}) 
and (\ref{6.29}) are gauge-invariant.
They can be re-written in another convenient form,
\be
P_L[q_\nu{}^\mu+N_\nu\gamma^\mu]\varphi_\mu\Big|_{\partial M}=0.
\ee
This defines eventually the boundary projector 
$P_{F}$ for the gauge operator $F$,
\be
P_F{}_\mu{}^\beta=[\delta_\mu{}^\nu-N_\mu N_\alpha\gamma^\alpha N^\nu
-N_\mu N_\alpha\gamma^\alpha\gamma^\nu]
P_{L}[q_\nu{}^\beta+N_\nu\gamma^\beta].
\ee
If the projector $P_L$ satisfies the symmetry 
condition (\ref{2.42aa}) then so does the 
projector $P_{F}$ (of course, one has to check it 
with the matrix $\Gamma^\mu N_\mu$).

Thus, we have shown that

\begin{theorem}
The boundary-value problem for the Rarita--Schwinger system 
with the boundary conditions (\ref{6.28a})--(\ref{6.29}) is gauge-invariant
and strongly elliptic provided that the projector $P_L$ satisfies 
the condition (\ref{2.42aa}).
Particular examples of such projectors are given 
by (\ref{2.57aab})--(\ref{2.57aa}).

\end{theorem}

\section{Euclidean quantum gravity}
\setcounter{equation}0

Generalized boundary 
conditions similar to the ones studied so far 
occur naturally in Euclidean quantum gravity \cite{barvi87,avra96}. 
The vector bundle $G$ is now the bundle of 
cotangent vectors, $G=T^*M $, and
$V$ is the vector bundle of symmetric rank-2
tensors (also called symmetric 2-forms) over $M$: $V=T^*M \vee T^*M $,
$\vee$ being the symmetrized tensor product.
The metric on the bundle $G$ is, naturally, 
just the metric on $M$, and
the fibre metric $E$ on the bundle $V$ 
is defined by the equation
\be
E^{ab\; cd}\equiv g^{a(c}g^{d)b}+\alpha g^{ab}g^{cd} ,
\label{5.1}
\ee
where $\alpha$ is a real parameter. One has also, for
the inverse metric,
\be
E^{-1}{}_{ab\; cd}\equiv g_{a(c}g_{d)b}
-{\alpha\over (1+\alpha m)}g_{ab}g_{cd} .
\ee
We do not fix the $\alpha$ parameter from the beginning, but rather
are going to study the dependence of the heat kernel on it.
It is not difficult to show that the 
eigenvalues of the matrix $E$ are $1$ (with degeneracy 
$m(m+1)/2-1$) and $(1+\alpha m)$. Therefore, 
this metric is positive-definite only for
\be
\alpha>-{1\over m},
\label{142f}
\ee
and becomes singular for $\alpha=-1/m$. Thus, hereafter
$\alpha\ne -1/m$.

The generator $R$ of infinitesimal gauge transformations 
is now defined to be the Lie derivative of the
tensor field $\varphi$ along the vector field $\epsilon$,
\be
(R\epsilon)_{ab} \equiv (L_{\epsilon}\varphi)_{ab}
={\sqrt 2}\nabla_{(a}\epsilon_{b)}.
\ee
The adjoint generator $\bar R$ with the metric 
$E$ is defined by its action,
\be
(\bar R\varphi)_{a} \equiv -{\sqrt 2}E_a{}^{b\,cd}\nabla_b\varphi_{cd}.
\ee
The leading symbol of the ghost operator 
$L=\bar R R$ takes now the form
\be
\sigma_L(L;\xi)\epsilon_a=2E_a{}^{bcd}\xi_b\xi_c\epsilon_d
=\left[\delta^b_a|\xi|^2+(1+2\alpha)\xi_a\xi^b\right]\epsilon_{b}.
\ee
Therefore, we see that it becomes of Laplace 
type only for $\alpha=-1/2$.
Note also that the operator $L$ has positive-definite leading symbol
only for $\alpha>-1$, and becomes degenerate for $\alpha=-1$.
Further, the leading symbol of the operator $H=R\bar R$ reads
\be
\sigma_L(H;\xi)\varphi_{ab}=
2\xi_{(a}E_{b)}{}^{cde}\xi_c\varphi_{de}
=\left[2\xi_{(a}\delta^{(c}_{b)}\xi^{d)}
+2\alpha\xi_a\xi_b g^{cd}
\right]\varphi_{cd}.
\ee

The gauge-invariant operator $\Delta$ is well 
known (see, e.g. \cite{dewitt83,avra91}). 
It has the following leading symbol:
\be
\sigma_{L}(\Delta;\xi)\varphi_{ab}=
\left\{
\delta^{(c}_{(a}\delta^{d)}_{b)}|\xi|^2
+\xi_a\xi_bg^{cd}
-2\xi_{(a}\delta^{(c}_{b)}\xi^{d)}
+{1+2\alpha\over (1+\alpha m)}g_{ab}(\xi^c\xi^d-g^{cd})
\right\}\varphi_{cd}.
\ee
Thus, we see that 
the operator $F=\Delta+H$ is of Laplace type only
in the case $\alpha=-1/2$. Let us, however, consider
for the time being a Laplace-type operator $F$ on symmetric
rank-2 tensors with a fibre metric (\ref{5.1}) with an 
{\it arbitrary} parameter $\alpha$.

Further, we define the projector $\Pi$ 
\be
\Pi=\Pi_{ab}{}^{cd} \equiv q_{(a}{}^c\; q_{b)}{}^d
-{\alpha\over (\alpha+1)}N_aN_bq^{cd} ,
\label{5.4}
\ee
where $q_{ab} \equiv g_{ab}-N_{a}N_{b}$.
It is not difficult to check that it is 
self-adjoint with respect to the metric $E$, i.e.
$\bar\Pi=\Pi$, and that
\be
\tr_{V}\Pi={1\over 2}m(m-1).
\label{136f}
\ee
Thus, we consider a Laplace-type operator $F$ 
acting on symmetric rank-2 tensors
with the following boundary conditions:
\be
\Pi_{ab}{}^{cd}\varphi_{cd}\Big|_{\partial M}= 0,
\label{555}
\ee
\be
E^{ab\, cd}\nabla_b \varphi_{cd}\Big|_{\partial M}=0 .
\label{5.2}
\ee

Separating the normal derivative we find
from here the boundary operator $B_F$ of the form 
(\ref{25e}), the operator $\Lambda$ being given by (\ref{55}) with
the matrices $\Gamma^i$ defined by
\be
\Gamma^{i} = \Gamma^i{}_{ab}{}^{cd} \equiv
 -{1\over (1+\alpha)}N_{a}N_{b}e^{i(c}N^{d)}
+N_{(a}e_{b)}^iN^{c}N^{d}.
\label{5.5a}
\ee
It is not difficult to check that these matrices are 
anti-self-adjoint and satisfy
the conditions (\ref{34e}) and (\ref{55d}).
The matrix ${T} \equiv {}\Gamma\cdot\zeta $ 
reads (cf. \cite{avresp97b}) 
\be
{T}=-{1\over (1+\alpha)}p_{1}+p_{2},
\ee
where
\be
p_{1} = p_1{}_{ab}{}^{cd} \equiv N_{a}N_{b}\zeta^{(c}N^{d)},
\qquad
p_{2} = p_2{}_{ab}{}^{cd} \equiv N_{(a}\zeta_{b)}N^cN^d,
\ee
and $\zeta_a=e^i_a\zeta_i$, so that $\zeta_{a} N^{a}=0$.
It is important to note that
\be
\Pi {T}={T}\Pi=0.
\ee
We also define further projectors,
\be
\rho = \rho_{ab}{}^{cd}
\equiv {2\over |\zeta|^2}N_{(a}\zeta_{b)}N^{(c}\zeta^{d)},
\label{144f}
\ee
\be
p = p_{ab}{}^{cd}
\equiv N_{a}N_{b}N^{c}N^{d},
\ee
which are mutually orthogonal:
\be
p\rho=\rho p=0.
\ee
The matrices $p_{1}$ and $p_{2}$, however, are nilpotent:
$p_1^2=p_2^2=0$, and their products are proportional 
to the projectors:
\be
p_{1}p_{2}={1\over 2}|\zeta|^2 p,\qquad
p_{2}p_{1}={1\over 2}|\zeta|^2\rho.
\ee
Therefore, we have
\be 
{T}^2=-{1\over 2(1+\alpha)}|\zeta|^{2}(p+\rho)
\equiv -\tau^2(p+\rho),
\label{148f}
\ee
where
\be
\tau \equiv {1\over\sqrt{2(1+\alpha)}}|\zeta|.
\label{146d}
\ee
We compute further
\be
{T}^{2n}=(i\tau)^{2n}(p+\rho),
\ee
\be
{T}^{2n+1}=(i{\tau})^{2n}{T}.
\ee
Last, by using 
\be
\tr p=\tr\rho=1,
\label{153f}
\ee
\be
\tr p_{1}=\tr p_{2}=0,
\ee
we obtain
\be
\tr {T}^{2n}=2(i\tau)^{2n},
\qquad
\tr {T}^{2n+1}=\tr {T}=0.
\ee

This suffices to prove:
\begin{lemma}
For any function $f$ analytic in the region 
$|z|\le \tau$, one has
\bea
f({T})&=&f(0)\left[\II-p-\rho\right]
+{1\over 2}[f(i\tau)+f(-i\tau)](p+\rho)
\nonumber\\
&&
+{1\over 2i\zeta}[f(i\tau)-f(-i\tau)]{T},
\label{157e}
\eea
\be
\tr f({T})= \left[{m(m+1)\over 2}-2\right]f(0)
+f(i\tau)+f(-i\tau).
\label{7.29nnny}
\ee
Thus, the eigenvalues of the matrix ${T}$ are
\be
{\rm spec}\, ({T})=
\left\{
\begin{array}{ll}
0 & {\rm with \ degeneracy }\quad
\left[{m(m+1)\over 2}-2\right]\\
i\tau & {\rm with \ degeneracy }\quad 1\\
-i\tau & {\rm with \ degeneracy }\quad 1
\end{array}
\right.
\ee
with $\tau$ defined by eq. (\ref{146d}).
\end{lemma}

This means that the eigenvalues of the matrix ${T}^2$ for a 
non-vanishing $\zeta_j$ are $0$ and $-1/[2(1+\alpha)]|\zeta|^2$.
Thus, the strong ellipticity condition (\ref{52eaaa}), which means that
the matrix $({T}^2+\zeta^{2}\II)$, for non-zero $\zeta$, should be 
positive-definite, takes the form
\be
-{1\over 2(1+\alpha)}+1>0.
\ee
This proves eventually 
\begin{theorem}
The boundary-value problem for a 
Laplace-type operator acting on sections of the bundle of symmetric
rank-2 tensors with the boundary conditions (\ref{555})
and (\ref{5.2}) is strongly elliptic with respect to
${\bf C}-{\bf R}_{+}$ only for
\be
\alpha>-{1\over 2}.
\label{150f}
\ee
\end{theorem}

\paragraph{Remarks.}
First, let us note that the condition (\ref{150f}) of strong 
ellipticity is compatible with the condition (\ref{142f})
of positivity of the fibre metric $E$. Second, it is exactly the
value $\alpha=-1/2$ that appears in the gauge-invariant
boundary conditions in one-loop quantum gravity 
in the minimal DeWitt gauge.
For a general value $\alpha\ne -1/2$, the operator $F$ 
resulting from $\Delta$ and $H$ is {\it not} of Laplace type, which
complicates the analysis significantly.
In other words, we have a corollary \cite{avresp97b}:

\begin{corollary}
The boundary-value problem in one-loop Euclidean 
quantum gravity, with a Laplace-type
operator $F$ acting on rank-two symmetric tensor fields, and with the 
gauge-invariant boundary conditions
(\ref{555}) and (\ref{5.2}), the 
fibre metric being $E$ with $\alpha=-1/2$,
is not strongly elliptic with respect to ${\bf C}-{\bf R}_{+}$.
\end{corollary}

We can also evaluate the coefficient $a_{1/2}$ of heat-kernel asymptotics.
This is most easily obtained by using the representation (\ref{165f})
of the coefficient $\beta_{1/2}$ in form of a Gaussian integral.
Using eq. (\ref{148f}) we get first
\be
\exp(-{T}^2)-\II=\left\{\exp\left[{1\over 2(1+\alpha)}
|\zeta|^2\right]-\II\right\}(p+\rho).
\ee
Therefore, from eq. (\ref{165f}) we obtain 
\be
{\beta_{1/2}}=2\int\limits_{{\bf R}^{m-1}}{d\zeta \over \pi^{(m-1)/2}}
\left\{\exp\left[-{1+2\alpha\over 2(1+\alpha)}|\zeta|^2
\right]-e^{-|\zeta|^2}
\right\}(p+\rho).
\ee
One should bear in mind that 
$\rho$ is a projector that depends on $\zeta$ (see (\ref{144f})).
Although $\rho$ is singular at the point $\zeta=0$, the integral is well
defined because of the difference of two exponential functions.
Calculating the Gaussian integrals we obtain
\be
\beta_{1/2}=2\left[\left({2(\alpha+1)\over 
1+2\alpha}\right)^{(m-1)/2}-1\right]
\left(p+{1\over (m-1)}\psi\right),
\ee
\be
a_{1/2}={\sqrt \pi\over 2}
\left\{\II-2\Pi+
2\left[\left({2(\alpha+1)\over 1+2\alpha}\right)^{(m-1)/2}-1\right]
\left(p+{1\over (m-1)}\psi\right)
\right\},
\ee
where $\psi$ is yet another projector:
\be
\psi = \psi_{ab}{}^{cd} \equiv 2N_{(a}q_{b)}^{\; (c}N^{d)}.
\label{444f}
\ee
Last, from eq. (\ref{104f}), by using the 
traces of the projectors (\ref{136f}),
(\ref{153f}) and (\ref{444f}), and the dimension 
of the bundle of symmetric rank-2 tensors,
\be
\dim V=\tr_V\II={1\over 2}m(m+1),
\ee
we obtain
\be
\tr_V{a_{1/2}}={\sqrt\pi\over 2}\left\{-{1\over 2}m(m-3)-4
+4\left[{2(\alpha+1)\over 1+2\alpha}\right]^{(m-1)/2}
\right\}.
\ee
Thus, we see that there is a singularity at $\alpha=-1/2$, which 
reflects the lack of strong ellipticity in this case.

\subsection{Heat-kernel diagonal in the non-elliptic case}

Consider now the case $\alpha=-1/2$.
{}From eq. (\ref{157e}) we have, in particular, 
\bea
(\II\,\mu-i{T})^{-1}
&=&{1\over \mu}(\II-p-\rho)
+{\mu\over \mu^2-|\zeta|^2}(p+\rho)
+i{1\over \mu^2-|\zeta|^2}{T}
\nonumber\\
&=&
{1\over \mu}(\II-p-\rho)
-{\mu\over \lambda}(p+\rho)
-i{1\over \lambda}{T},
\label{106}
\eea
\be
\det\left(\II\,\mu-i{T}\right)=\mu^{m(m+1)/2-2}(\mu^2-|\zeta|^2)
=(|\zeta|^2-\lambda)^{m(m+1)/4-1}(-\lambda).
\label{105}
\ee
For the boundary-value problem to be strongly elliptic, 
this determinant should be non-vanishing
for any $(\zeta,\lambda)\ne (0,0)$ and $\lambda\in 
{\bf C}-{\bf R}_+$, including the 
case $\lambda=0$, $\zeta\ne 0$.
Actually we see that, for any non-zero 
$\lambda\notin  {\bf R}_+$, this determinant
does not vanish for any $\zeta$. However, 
for $\lambda=0$ and any $\zeta$ it equals zero,
which means that the corresponding boundary conditions {\it do not} fix 
a unique solution of the eigenvalue equation for the leading 
symbol, subject to a decay condition at infinity. 
This is reflected by the simple fact that the coefficient ${a_{1/2}}$ 
of the asymptotic expansion of the heat kernel is not well defined, 
in that the integrals that determine it are divergent.
At the technical level, the non-ellipticity is reflected 
in the fact that the heat-kernel diagonal,
although well defined, has a {\it non-standard non-integrable} 
behaviour near the boundary, i.e. for $r \to 0$.

To prove this property,
let us calculate the fibre trace of the heat-kernel diagonal.
{}From (\ref{152pp}) we have
\be
\tr_V U_0(t|x,x)=(4\pi t)^{-m/2}\left[c_0
+c_1e^{-r^2/t}+J(r/\sqrt t)\right],
\label{152c}
\ee
where
\be
c_{0} \equiv \dim V={m(m+1)\over 2},
\ee
\be
c_{1} \equiv \tr_V(\II-2\Pi)=-{m(m-3)\over 2},
\ee
\be
J(z) \equiv \tr_V \Phi(z)=
-2\int\limits_{{\bf R}^{m-1}} {d\zeta\,\over \pi^{(m-1)/2}}\,
\int\limits_C{d\omega\over\sqrt\pi}\, 
e^{-|\zeta|^2-\omega^2+2i\omega z}\,
\tr_V
\Gamma\cdot\zeta
(\omega\,\II+\Gamma\cdot\zeta)^{-1}.
\ee
Now, for $\alpha=-1/2$
the parameter $\tau$ (\ref{146d}) determining the 
eigenvalues of the matrix $T=\Gamma\cdot\zeta$ is equal to  $\tau=|\zeta|$.
By using (\ref{7.29nnny}) we get 
\be
J(z)=
-4\int\limits_{{\bf R}^{m-1}} {d\zeta\,\over \pi^{(m-1)/2}}\,
\int\limits_C{d\omega\over\sqrt\pi}\, 
e^{-|\zeta|^2-\omega^2+2i\omega z}\,
{|\zeta|^2\over \omega^2+|\zeta|^2}
\ee
Remember that the contour $C$ lies in the upper 
half plane: it comes from
$-\infty+i\varepsilon$, encircles the point 
$\omega=i|\zeta|$ in the clockwise
direction and goes to $\infty+i\varepsilon$. The integral over $\omega$ 
is calculated by using the formula
\be
\int\limits_C d\omega\, f(\omega)=
-2\pi i\, {\rm Res}_{\omega=i|\zeta|}\, f(\omega)
+
\int\limits_{-\infty}^{\infty}
d\omega\,f(\omega).
\ee
The integrals over $\zeta$ can be reduced to Gaussian integrals by 
lifting the denominator in the exponent (cf. (\ref{383})) or by
using spherical coordinates. In this way we get eventually
\be
J(z)= 2(m-1)\Gamma(m/2) \,{1\over z^{m}}
-2(m-1)\int\limits_{0}^{1}du\,u^{m/2-1}e^{-z^2 u}.
\ee
This can also be written in the form
\be
J(z)= 2(m-1)z^{-m}\left[\Gamma(m/2)-\gamma(m/2,z^{2})\right]
=2(m-1)z^{-m}\Gamma(m/2,z^{2}),
\label{168e}
\ee
by using the incomplete $\gamma$-functions
\be
\gamma(a,x) \equiv \int\limits_{0}^{x}
du\, u^{a-1}e^{-u}, \qquad
\Gamma(a,x) \equiv \int_{x}^{\infty}du \, 
u^{a-1}e^{-u}.
\ee
It is immediately seen that the function $J(z)$ is singular
as $z\to 0$
\be
J(z)\sim 2(m-1)\Gamma(m/2) \,{1\over z^{m}}.
\ee
By using the identity
\be
\Gamma(a,x)=x^{a-1}e^{-x}+(a-1)\Gamma(a-1,x)
\ee
we also find that, when $z \rightarrow \infty$, $J(z)$ is
exponentially small, i.e.
\be
J(z) \sim 2(m-1)z^{-2}e^{-z^{2}}.
\label{235aaaz}
\ee
Note that, for $\alpha=-{1\over 2}$, one has $\tr_{V}T^{2}
=-2 {|\zeta|}^{2}$, and hence $\tr_{V}\Gamma^{i}\Gamma^{j}
=-2g^{ij}$ and $\tr_{V}\Gamma^{2}=-2(m-1)$. Thus, we see that
the asymptotics (\ref{235aaaz}) as $z \rightarrow \infty$
corresponds to eq. (\ref{3.87aaaz}).

The singularity at the point $z=0$ results exactly from the 
pole at $\omega=i |\zeta|$. In the strongly elliptic case
all poles lie on the positive imaginary line with 
${\rm Im}\,\omega <i|\zeta|$, so that there is a {\it finite} 
gap between the pole located at the point with the largest value
of the imaginary part and the point $i|\zeta|$.

Thus, we obtain
\bea
\tr_V U_0(t|x,x)=(4\pi t)^{-m/2}\left 
[c_{0}+c_{1}e^{-{r^2/t}}
+2(m-1)\left({r\over\sqrt t}\right)^{-m}
\Gamma(m/2,r^2/t) \right].
\label{232e}
\eea
Here the first term is the standard first term in 
the heat-kernel asymptotics which
gives the familiar interior contribution when 
integrated over a compact manifold. The boundary terms in 
(\ref{232e}) are exponentially small as $t \rightarrow 0^{+}$,
if $r$ is kept fixed. Actually, these terms should behave, as
$t \rightarrow 0^{+}$, as distributions near the boundary, so that
they give well defined non-vanishing contributions (in form of
integrals over the boundary) when integrated with a smooth 
function. The third term, however, gives rise to an unusual 
singularity at the boundary as $r \rightarrow 0$ 
(on fixing $t$):
\be
\tr_VU_{0}(t|x,x)\stackrel{r\to 0}{\sim}(4\pi)^{-m/2}
2(m-1)\Gamma(m/2) {1\over r^{m}}.
\ee
This limit is {\it non-standard} in that: i) it does not
depend on $t$ and ii) it is {\it not integrable} over $r$ 
near the boundary, as $r \rightarrow 0$.
This is a direct consequence of the lack of strong ellipticity.

\section{Concluding remarks}
\setcounter{equation}0

We have studied a generalized boundary-value problem for operators of 
Laplace type, where a part of the field is subject to Dirichlet
boundary conditions, and the remaining part is subject to
conditions which generalize the Robin case by the inclusion 
of tangential derivatives. The corresponding boundary operator
can be always expressed in the form 
(\ref{25e}), where $\Lambda$ is a tangential
differential operator of the form (\ref{55}). The fermionic
analysis for Dirac-type operators has also been developed.
The strong ellipticity of the resulting boundary-value 
problem is crucial, in particular,
to ensure the existence of the asymptotic expansions frequently
studied in the theory of heat kernels \cite{gilkey95}. In
physical problems, this means that the one-loop semi-classical
expansions of the Green functions and of the effective action
in quantum field theory are well defined and can 
be computed explicitly on compact 
Riemannian manifolds with smooth boundary. The occurrence of boundaries
plays indeed a key role in the path-integral approach to quantum
gravity \cite{hawk79}, and it appears desirable to study the strong
ellipticity problem for all gauge theories of physical interest,
now that a unified scheme for the derivation of BRST-invariant
boundary conditions is available \cite{moss97}. 

We have thus
tried to understand whether the following requirements are
compatible:
\begin{itemize}
\item[(i)]
The gauge-field operator, $F$, should be of Laplace type
(and the same for the ghost operator);
\item[(ii)]
Local nature of the boundary operator $B_{F}$ (in 
fact, we have studied the case
when $B_F$ is a first-order differential operator, and
boundary projectors for fermionic fields);
\item[(iii)]
Gauge invariance of the boundary conditions;
\item[(iv)]
Strong ellipticity of the boundary-value problem
$(F,B_{F})$.
\end{itemize}
First, we have found a condition of strong ellipticity  
(see (\ref{52eaa})) for a generalized boundary-value problem
for a Laplace-type operator. For operators of Dirac type, one
finds instead the strong ellipticity properties described in our
Theorem 4. Second, we have constructed the resolvent kernel and the
heat kernel in the leading approximation
(sect. 3.1 and sect. 3.2) and evaluated the first 
non-trivial heat-kernel coefficient 
$A_{1/2}$  in Eqs. (\ref{3.45ee}) and (\ref{3.49ee}). Third, we 
have found a criterion of strong ellipticity of a
general gauge theory in terms of the gauge generators 
(see (\ref{203j})). As physical applications
of the above results we have studied the Yang--Mills, Rarita--Schwinger
and Einstein field theories. Interestingly, only in the latter the
strong ellipticity condition is not satisfied if the
conditions (i), (ii) and (iii) hold. Moreover, the gauge-invariant
boundary conditions for the Rarita--Schwinger system have been
found to involve only the boundary projector for the ghost operator.
As far as we know, our results as well as the consequent 
analysis of the physical gauge models, are completely original, or extend
significantly previous work in the literature
\cite{mcavity91,avresp97b,dowker97}. 

Since we find that, for gravitation,
the four conditions listed above are not,
in general, compatible, it seems that one should
investigate in detail at least one of the following alternatives:

\begin{itemize}
\item[(1)] 
Quantum gravity and quantum supergravity on manifolds with
boundary with gauge-field operators which are non-minimal;
\item[(2)]
Non-local boundary conditions for gravitation \cite{mara96}
and spin-3/2 fields \cite{deat91b};
\item[(3)]
Non-gauge-invariant `regularization' of the boundary conditions, which 
suppresses the tangential derivatives and improves the ellipticity;
\item[(4)]
Boundary conditions which are not completely gauge-invariant,
and hence avoid the occurrence of tangential derivatives in the
boundary operator.
\end{itemize}

The latter possibility has been investigated in 
\cite{luck91} and has been widely
used in the physical literature (see \cite{espo97} and
references therein). The first 3 options are not yet (completely)
exploited in the literature. In particular, it is unclear
whether one has to resort to non-minimal operators to preserve
strong ellipticity of the boundary-value problem. Moreover, 
non-local boundary conditions which are completely gauge-invariant,
compatible with local supersymmetry transformations at the
boundary (cf. \cite{luck91}), and ensure strong ellipticity, are not
easily obtained (if at all admissible). Indeed, it should
not be especially surprising that gauge theories have, in general, an
{\it essentially non-local} character. Since the projectors on the physical
gauge-invariant subspace of the configuration space are non-local,
a consistent formulation of gauge theories on Riemannian manifolds 
(even without boundary) is {\it necessarily} non-local (see \cite{avra91}).  

Thus, there is increasing
evidence in favor of boundaries raising deep 
and unavoidable foundational issues
for the understanding of modern quantum field theories
\cite{vasil97}. The
solution of such problems might in turn shed new light on
spectral geometry and on the general theory of elliptic operators
\cite{gilkey95,espo97b}.

\section*{Acknowledgements}
We are much indebted to Andrei Barvinsky, 
Thomas Branson, Stuart Dowker, Peter Gilkey, Alexander
Kamenshchik, Klaus Kirsten and Hugh Osborn for
correspondence and discussion of the generalized
boundary conditions, and to Stephen Fulling for valuable remarks.
The work of IA was
supported by the Deutsche Forschungsgemeinschaft.

\end{document}